\documentclass{style-class/pasa}%

\usepackage{graphicx}
\usepackage{mathtools}
\usepackage{multirow}
\usepackage{bm}
\usepackage{textcomp}

\newcommand{\mathbfss}[1]{\textbf{\textsf{#1}}}
\newcommand{\eppsilon}{$\varepsilon$ppsilon}
\DeclarePairedDelimiter{\diagpars}{(}{)}
\newcommand{\diag}{\operatorname{diag}\diagpars}

\title[FHD/{\eppsilon} EoR PS Pipeline]{The FHD/$\bm{\varepsilon}$ppsilon Epoch of Reionization Power Spectrum Pipeline}

\author[N. Barry et al.]{N. Barry$^{1,2}$\thanks{Contact: nichole.barry@unimelb.edu.au}, A. P. Beardsley$^3$, R. Byrne$^{4}$, B. Hazelton$^{4,5}$, M. F. Morales$^{4,6}$, J. C. Pober$^{7}$, and I. Sullivan$^{8}$
\affil{$^1$School of Physics, The University of Melbourne, Parkville, VIC 3010, Australia}
\affil{$^2$ARC Centre of Excellence for All Sky Astrophysics in 3 Dimensions (ASTRO 3D)}
\affil{$^3$School of Earth and Space Exploration, Arizona State University, Tempe, AZ 85287, USA}
\affil{$^4$Department of Physics, University of Washington, Seattle, WA 98195, USA}
\affil{$^5$eScience Institute, University of Washington, Seattle, WA 98195, USA}
\affil{$^6$Dark Universe Science Center, University of Washington,
Seattle, 98195, USA}
\affil{$^7$Department of Physics, Brown University, Providence, RI 02906, USA}
\affil{$^8$Department of Astronomy, University of Washington, Seattle, WA 98195, USA}
}%

\jid{PASA}
\doi{10.1017/pas.\the\year.xxx}
\jyear{\the\year}

\usepackage{style-class/aas_macros}
\usepackage{hyperref} 
\hypersetup{colorlinks,citecolor=blue,linkcolor=blue,urlcolor=blue}

\begin{document}

\begin{frontmatter}
\maketitle

\begin{abstract}
Epoch of Reionization data analysis requires unprecedented levels of accuracy in radio interferometer pipelines. We have developed an imaging power spectrum analysis to meet these requirements and generate robust 21\,cm EoR measurements. In this work, we build a signal path framework to mathematically describe  each step in the analysis, from data reduction in the FHD package to power spectrum generation in the {\eppsilon} package. In particular, we focus on the distinguishing characteristics of FHD/{\eppsilon}: highly accurate spectral calibration, extensive data verification products, and end-to-end error propagation. We present our key data analysis products in detail to facilitate understanding of the prominent systematics in image-based power spectrum analyses. As a verification to our analysis, we also highlight a full-pipeline analysis simulation to demonstrate signal preservation and lack of signal loss. This careful treatment ensures that the FHD/{\eppsilon} power spectrum pipeline can reduce radio interferometric data to produce credible 21\,cm EoR measurements.

\end{abstract}

\begin{keywords}
cosmology: dark ages, reionization, first stars -- techniques: interferometric -- methods: data analysis -- instrumentation: interferometers
\end{keywords}
\end{frontmatter}


\section{Introduction}
\label{sec:intro}

Structure measurements of the Epoch of Reionization (EoR) have the potential to revolutionize our understanding of the early universe. Many radio interferometers are pursuing these detections, including the MWA \citep{tingay_murchison_2013,bowman_science_2013,wayth_phase_2018}, PAPER \citep{parsons_precision_2010}, LOFAR \citep{yatawatta_initial_2013,van_haarlem_lofar:_2013}, HERA \citep{pober_what_2014,deboer_hydrogen_2017}, and the SKA \citep{mellema_reionization_2013,koopmans_cosmic_2015}. The sheer amount of data to be processed even for the most basic of interferometric analyses requires sophisticated pipelines.

Furthermore, these analyses must be very accurate; the EoR signal is many orders of magnitude fainter than the foregrounds. Systematics from inaccurate calibration, spatial/frequency transform artifacts, and other sources of spectral contamination will preclude an EoR measurement. Understanding and correcting for these systematics has been the main focus of improvements for MWA Phase I EoR analyses, including the Fast Holographic Deconvolution (FHD)/{\eppsilon} pipeline, since their description in \citet{jacobs_murchison_2016}.

FHD/{\eppsilon} power spectrum analyses have been prevalent in the literature, including data reduction for MWA Phase I \citep{beardsley_first_2016}, MWA Phase II \citep{li_comparing_2018}, and for PAPER \citep{kerrigan_improved_2018}, with HERA analyses planned. It is also flexible enough to be used for pure simulation, particularly in investigating calibration effects \citep{barry_calibration_2016,byrne_fundamental_2019}. Combined with its focus on spectrally accurate calibration, as well as end-to-end data-matched model simulations and error propagation, FHD/{\eppsilon} is well-suited for EoR measurements and studies.

In this work, we describe the FHD/{\eppsilon} power spectrum pipeline in full with a consistent mathematical framework, highlighting recent improvements and key features. We trace the signal path from its origin through the main components of our data reduction analysis. Our final power spectrum products, including uncertainty estimates, are detailed extensively. We perform many diagnostics with various types of power spectra, which provide confirmations for our improvements to the pipeline.

As further verification of the FHD/{\eppsilon} power spectrum pipeline, we present proof of signal preservation. Confidence in EoR upper limits relies on our ability to avoid absorption of the signal itself. A full end-to-end analysis simulation within our pipeline proves that we do not suffer from signal loss, thereby adding credibility to our EoR upper limits.

First, we detail the signal path framework in \S\ref{sec:signal_path} to provide a mathematical foundation. In \S\ref{sec:FHD}, we build an analytical description of FHD, focusing on the recent improvements in calibration. \S\ref{sec:integration} outlines the data products from FHD and our choice of integration methodology. Propagating errors and creating power spectra in {\eppsilon} is described in \S\ref{sec:eppsilon}. Finally, we illustrate all of our power spectra data products in \S\ref{sec:power_spectrum_diagnostics}, including proof of signal preservation.

\section{The signal path framework}
\label{sec:signal_path}

The sky signal is modified during the journey from when it was emitted to when it was recorded. In order to uncover the EoR, the true sky must be separated from all intervening effects, including those introduced via the instrument. We describe all known signal modifications to build a consistent, mathematical framework.

\begin{figure}[t]
\centering
	\includegraphics[width = \columnwidth]{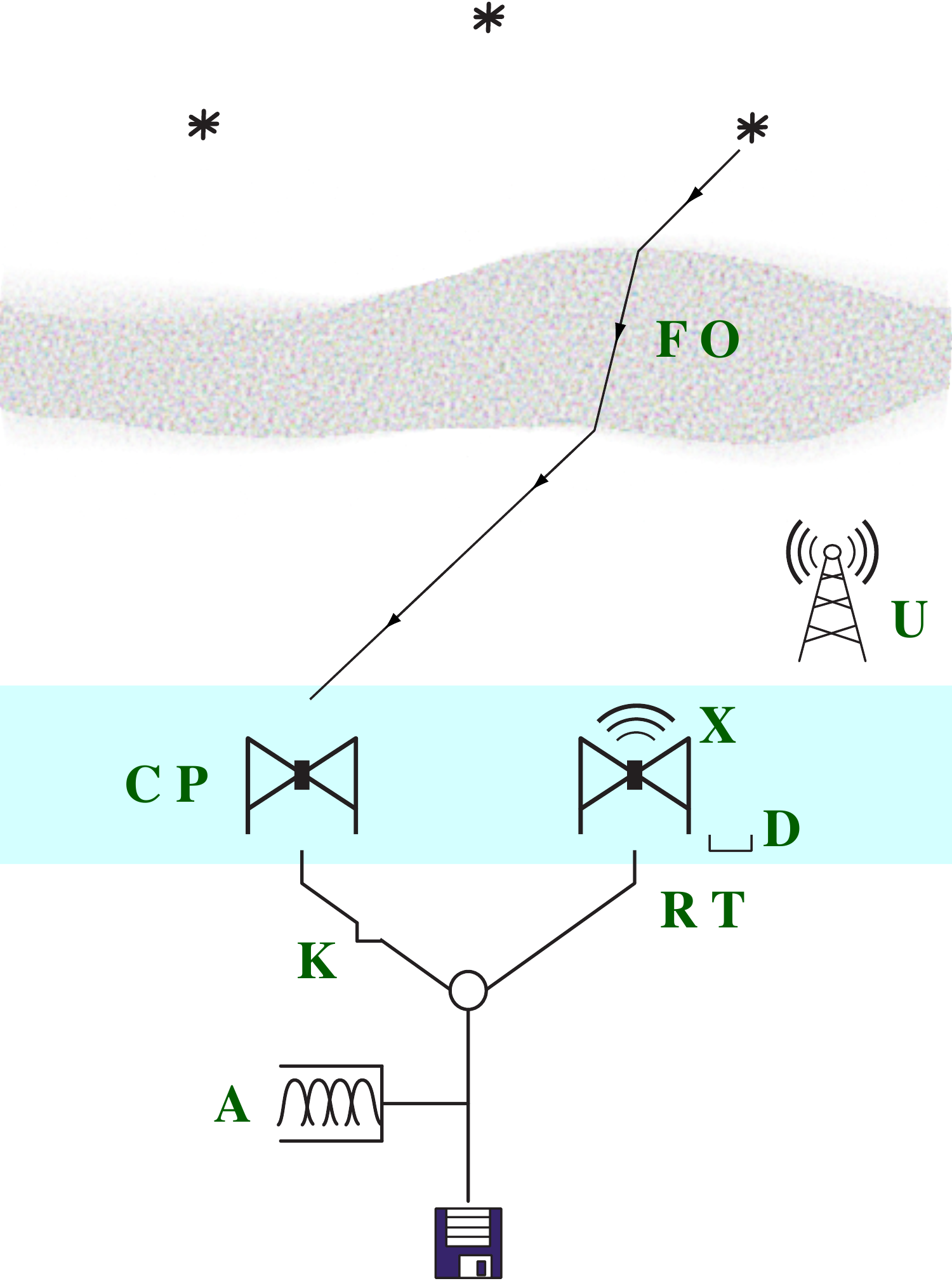}
	\caption[Signal path schematic]{The signal path through the instrument. There are three categories of signal modification: before antenna, at antenna (coloured light blue), and after antenna. Each modification matrix (green) is detailed in the text and Table~\ref{table:tablesignal}.}
	\label{fig:signal_path}
\end{figure}

There are three main types of signal modification: those that occur before, during, and after interaction with the antenna elements, as shown in Figure~\ref{fig:signal_path}. We adhere to notation from \citet{hamaker_understanding_1996} whenever possible in our brief catalog of interactions.

\begin{description}
    \item $\mathbfss{B}$: Before antenna
    \begin{itemize}
        \item Faraday rotation from interaction with the ionosphere, $\mathbfss{F}$.
        \item Source position offsets $\mathbfss{O}$ due to variation in ionospheric thickness.
        \item Unmodelled signals caused by radio frequency interference (RFI), $\mathbfss{U}$.
    \end{itemize}
    \item $\mathbfss{S}$: At antenna
    \begin{itemize}
        \item Parallactic rotation $\mathbfss{P}$ between the rotating basis of the sky and the basis of the antenna elements.
        \item Antenna correlations from cross-talk between antennas, $\mathbfss{X}$.
        \item Antenna element response, $\mathbfss{C}$, usually referred to as the beam\footnote{\citet{hamaker_understanding_1996} use $\mathbfss{C}$ to primarily account for a rotating feed, however we incorporate the antenna response as well. We also expect and model polarisation correlation in the Jones matrices (\S\ref{subsec:beam}), whereas \citet{hamaker_understanding_1996} does not.}.
        \item Errors in the expected nominal configuration and beam model, $\mathbfss{D}$.
    \end{itemize}
    \item $\mathbfss{E}$: After antenna
    \begin{itemize}
        \item Electronic gain amplitude and phase $\mathbfss{R}$ from a typical response of each antenna.
        \item Gain amplitude changes from temperature effects $\mathbfss{T}$ on amplifiers.
        \item Gain amplitude and phase oscillations $\mathbfss{K}$ due to cable reflections, both at the end of the cables and at locations where the cable is kinked.
        \item Frequency correlations $\mathbfss{A}$ caused by aliasing in polyphase filter banks or other channelisers.
    \end{itemize}
\end{description}

\begin{table}[tbh!]
\centering
    \begin{tabular}{c c c}
            Type & Variables & Definition \\
        \hline
            \multicolumn{1}{||c|}{\multirow{3}{*}{\begin{tabular}{@{}c@{}}$\mathbfss{B}$\end{tabular}}} &
            \multicolumn{1}{c|}{$\mathbfss{F}$} &
            \multicolumn{1}{l||}{Faraday rotation}   \\
        \cline{2-3}
            \multicolumn{1}{||c|}{} &
            \multicolumn{1}{c|}{$\mathbfss{O}$} &
            \multicolumn{1}{l||}{Source position offsets}  \\
        \cline{2-3}
            \multicolumn{1}{||c|}{} &
            \multicolumn{1}{c|}{$\mathbfss{U}$} &
            \multicolumn{1}{l||}{\begin{tabular}{@{}l@{}}Unmodelled RFI\end{tabular}}  \\
        \hline
        \hline
            \multicolumn{1}{||c|}{\multirow{4}{*}{$\mathbfss{S}$}} &
            \multicolumn{1}{c|}{$\mathbfss{P}$} &
            \multicolumn{1}{l||}{\begin{tabular}{@{}l@{}}Parallactic rotation \end{tabular}}\\
        \cline{2-3}
            \multicolumn{1}{||c|}{} &
            \multicolumn{1}{c|}{$\mathbfss{X}$} &
            \multicolumn{1}{l||}{\begin{tabular}{@{}l@{}} Cross-talk antenna correlations \end{tabular}}\\
        \cline{2-3}
            \multicolumn{1}{||c|}{} &
            \multicolumn{1}{c|}{$\mathbfss{C}$} &
            \multicolumn{1}{l||}{\begin{tabular}{@{}l@{}} Antenna element response\end{tabular}}\\
        \cline{2-3}
            \multicolumn{1}{||c|}{} &
            \multicolumn{1}{c|}{$\mathbfss{D}$} &
            \multicolumn{1}{l||}{\begin{tabular}{@{}l@{}} Errors in nominal configuration \end{tabular}}\\
        \hline
        \hline
            \multicolumn{1}{||c|}{\multirow{4}{*}{$\mathbfss{E}$}} &
            \multicolumn{1}{c|}{$\mathbfss{R}$} &
            \multicolumn{1}{l||}{\begin{tabular}{@{}l@{}}  Gain amplitude and phase\end{tabular}}\\
        \cline{2-3}
            \multicolumn{1}{||c|}{} &
            \multicolumn{1}{c|}{$\mathbfss{T}$} &
            \multicolumn{1}{l||}{\begin{tabular}{@{}l@{}} Temperature changes \end{tabular}}\\
        \cline{2-3}
            \multicolumn{1}{||c|}{} &
            \multicolumn{1}{c|}{$\mathbfss{K}$} &
            \multicolumn{1}{l||}{\begin{tabular}{@{}l@{}} Cable reflections\end{tabular}}\\
        \cline{2-3}
            \multicolumn{1}{||c|}{} &
            \multicolumn{1}{c|}{$\mathbfss{A}$} &
            \multicolumn{1}{l||}{\begin{tabular}{@{}l@{}} Frequency correlations \end{tabular}}\\
        \hline
        \hline
            \multicolumn{1}{||c|}{\multirow{3}{*}{$\textbf{v}$}} &
            \multicolumn{1}{c|}{$\textbf{I}^{\textrm{true}}$} &
            \multicolumn{1}{l||}{\begin{tabular}{@{}l@{}}  True sky\end{tabular}}\\
        \cline{2-3}
            \multicolumn{1}{||c|}{} &
            \multicolumn{1}{c|}{$\textbf{v}^{\textrm{meas}}$} &
            \multicolumn{1}{l||}{\begin{tabular}{@{}l@{}} Measured visibilities \end{tabular}}\\
        \cline{2-3}
            \multicolumn{1}{||c|}{} &
            \multicolumn{1}{c|}{$\textrm{noise}$} &
            \multicolumn{1}{l||}{\begin{tabular}{@{}l@{}} Thermal noise\end{tabular}}\\
        \hline

    \end{tabular}
    \caption{Brief definitions of the variables used within the signal path framework, organized by type. There are four types, $\mathbfss{B}$) interactions that occur before the antenna elements, $\mathbfss{S}$) interactions that occur at the antenna elements, $\mathbfss{E}$) interactions that occur after the antenna elements, and $\textbf{v}$) visibility-related variables. }
    \label{table:tablesignal}
\end{table}

Each contribution can be modelled as a matrix which depends on $[f,t,P,AB]$: frequency, time, instrumental polarisation, and antenna cross-correlations. The expected contributions  along the signal path are thus $\mathbfss{B} = \mathbfss{U}\mathbfss{O}\mathbfss{F}$, $\mathbfss{S} = \mathbfss{D}\mathbfss{C}\mathbfss{X}\mathbfss{P}$, and $\mathbfss{E} = \mathbfss{A}\mathbfss{K}\mathbfss{T}\mathbfss{R}$. The visibility measurement equation takes the form
\begin{equation}
    \textbf{v}^{\textrm{meas}} \sim \mathbfss{E} \ \mathbfss{S} \ \mathbfss{B} \  \textbf{I}^{\textrm{true}} + \textrm{noise},
    \label{eq:ISG}
\end{equation}
where $\textbf{v}^{\textrm{meas}}$ are the measured visibilities, $\textbf{I}^{\textrm{true}}$ is the true sky, $\mathbfss{B}$ are contributions that occur between emission and the ground, $\mathbfss{S}$ are contributions from the antenna configuration, and $\mathbfss{E}$ are contributions from the electronic response. We have included all known contributions and modifications to the signal, including the total thermal noise. However, there may be unknown contributions, hence we describe Equation~\ref{eq:ISG} as an approximation. All components are summarised in Table~\ref{table:tablesignal} for reference.

The visibilities can be condensed into vectors since the measurements are naturally independent over the $[f,t,P,AB]$ dimensions. However, the modification matrices can introduce correlations across the dimensions, and thus cannot be reduced without assumptions.

One of the goals of an imaging EoR analysis is to reconstruct the true sky visibilities, $\textbf{v}^{\textrm{true}}$, given the measured sky visibilities, $\textbf{v}^{\textrm{meas}}$. This is achieved through the process of calibration. We will use our generalized framework described in this subsection to detail the assumptions and methodology for the FHD/{\eppsilon} pipeline.

\section{Fast Holographic Deconvolution}
\label{sec:FHD}

FHD\footnote{\url{https://github.com/EoRImaging/FHD}} is an open-source radio analysis package that produces calibrated sky maps from measured visibilities. Initially built to implement an efficient deconvolution algorithm \citep{sullivan_fast_2012}, its purpose is now to serve as an vital step in EoR power spectrum analysis\footnote{Deconvolution in FHD is generally only used in building sky models for calibration and subtraction.}. 

FHD is a tool to analyse radio interferometric data, and has a series of main functions at its core: 1) creating model visibilities from sky catalogues for calibration and subtraction, 2) gridding calibrated data, and 3) making images for analysis and integration.

Using the signal path framework described in \S\ref{sec:signal_path}, we will describe the steps in building a gridding kernel (\S\ref{subsec:beam}), forming model visibilities (\S\ref{subsec:model_visibilities}), calibrating data visibilities (\S\ref{subsec:calibration}), and producing images (\S\ref{subsec:images}).

\subsection{Pre-pipeline flagging}
\label{subsec:flagging}

Before any analysis can begin, the data must be RFI-flagged. Radio frequency interference (RFI), particularly from FM radio and digital TV, can contaminate the data. However, RFI has characteristic signatures in time and frequency which allow it to be systematically removed by trained packages.

We use the package \textsc{aoflagger}\footnote{\url{https://sourceforge.net/p/aoflagger/wiki/Home/}} to RFI-flag the data \citep{offringa_low-frequency_2015}. This removes bright line-like emission, but has difficulty removing faint, broad emission like TV. We completely remove any observations that have signatures of TV. Therefore, we remove contributions from $\mathbfss{U}$ by avoidance.

As demonstrated in \citet{offringa_impact_2019}, averaging over flagged channels can cause bias. However, this does not affect the nominal FHD/{\eppsilon} pipeline because it avoids inverse-variance weighting of the visibilities. Any future incorporation of inverse-variance weighting will need to take this into account.

\subsection{Generating the beam}
\label{subsec:beam}

The measurement collecting area of an antenna element, $\mathbfss{C}$, is commonly referred to as the primary beam. A deep knowledge of the beam is critical for precision measurements with widefield interferometers \citep{pober_importance_2016}. Since our visibility measurements are correlations of the voltage response between elements, we must understand the footprint of each element's voltage response to reconstruct images \citep{morales_software_2009}.

We build the antenna element response from finely interpolated beam images. This happens in the instrument's coherency domain, or instrumental polarisation. We assume each element has two physical components $p$ and $q$ which are orthogonal, so each visibility correlation between elements $a$ and $b$ will have a $P=\{p_ap_b, p_aq_b, q_ap_b, q_aq_b\}$ response in the coherency domain \citep{hamaker_understanding_1996}. This is calculated from the two elements' polarised response patterns. For example,
\begin{multline}
    \mathbfss{C}_{p_aq_b} = \mathbfss{C}_{q_ap_b} = \\ \bigg(\left(\mathbfss{J}_{x,p_a}^{\phantom{*}} \mathbfss{J}_{x,p_b}^* + \mathbfss{J}_{y,p_a}^{\phantom{*}} \mathbfss{J}_{y,p_b}^*\right) \\ \left(\mathbfss{J}_{x,q_a}^{\phantom{*}} \mathbfss{J}_{x,q_b}^* + \mathbfss{J}_{y,q_a}^{\phantom{*}} \mathbfss{J}_{y,q_b}^*\right)\bigg)^{\frac{1}{2}},
\end{multline}
where $\mathbfss{C}_{p_aq_b}$ is the beam response for a $p$ component in element $a$ and a $q$ component in element $b$, $\mathbfss{J}$ is the vector field for an element (also known as the Jones matrix), and the subscript $x,p_a$ is the contribution of the $p$ component in element $a$ for the coordinate $x$ (and likewise for the other subscripts) \citep{sutinjo_understanding_2015}. Each matrix is a function of spatial coordinates and frequency, and each operation is done element-by-element.

The Jones matrices describe the transformation needed to account for $\mathbfss{P}$, or parallactic rotation. Due to the wide field-of-view and the lack of moving parts for most EoR instruments, this is a natural requirement. Known inter-dipole mutual coupling, a form of $\mathbfss{X}$, is also captured in the Jones matrices. We generate $\mathbfss{C}_{P}$ for a pair of elements, which can be applied to all other identical element pairs.

We then take the various beams in direction cosine space $\{l,m\}$ and Fourier transform them to get beams in $\{u,v\}$. Since we plan on using the beam as a gridding kernel later on, the beam must vary as smoothly as possible. This is achievable by hyperresolving beyond the usual $uv$-grid resolution of $\frac{1}{2}\lambda$, down to typically $\frac{1}{7000}\lambda$. We create this beam once to build a highly-resolved reference table.

FHD does have the flexibility to generate unique beams for each element given individual element metadata, however this quickly increases computing resources. Instead, we can build a coarse beam per baseline with phase offsets in image space to account for pixel center offsets \citep{line_puma_2017}, which is built on-demand rather than saved as a reference table. This corrects for one form of $\mathbfss{D}$, or individual element variation and error.

When we generate model visibilities, we want to be as instrumentally accurate as possible. This necessitates a kernel which represents the instrumental response to the best of our knowledge. However, we can choose a separate, modified kernel for $uv$-plane generation in power spectrum analysis, similar to a Tapered Gridded Estimator \citep{choudhuri_visibility-based_2014, choudhuri_tapering_2016}. As long as the proper normalisations are taken into account in power estimation, a modified gridding kernel acts as a weighting of the instrumental response. This image-space weighting will correlate pixels; we investigate contributions from pixel correlations in the power spectrum in \S\ref{subsec:2D}.

\subsection{Creating model visibilities}
\label{subsec:model_visibilities}

We must calibrate our input visibilities. Due to our wide field-of-view and accuracy requirements, FHD simulates \textit{all} reliable sources out to typically 1\% beam level in the primary lobe and the sidelobes to build a nearly complete theoretical sky. For the MWA Phase I, the 1\% beam level includes the first sidelobe out to a field-of-view of approximately 100\textdegree. By comparing these model visibilities to the data visibilities, we can estimate the instrument's contribution.

Generating accurate model visibilities is an important step in calibration. Therefore, the estimate of the sky must be as complete as possible, including source positions, morphology, and brightness. For example, we can use GLEAM \citep{hurley-walker_GLEAM}, an extragalactic catalogue with large coverage and high completeness, to model sources for observations in the Southern sky. For a typical MWA field off the galactic plane, we model over 50,000 known sources using GLEAM. As for a typical PAPER observation, we model over 10,000 known sources above 1\,Jy \citep{kerrigan_improved_2018}.

Not all sources are unpolarised. There are cases where a source can be linearly or circularly polarized. Only ten out of the thousands of sources seen in a typical MWA field off the galactic plane are known to be reliably polarised \citep{riseley_pogs_2018}. Therefore, we assume there are no polarised sources in making model visibilities. With this assumption, we avoid complications from $\mathbfss{F}$, or Faraday rotation in the polarisation components due to the ionosphere. Implicitly, we also assume the ionosphere is not structured, therefore $\mathbfss{F}$ does not affect unpolarised sources.

We disregard source position offsets $\mathbfss{O}$, a contamination resulting from ionospheric distortions. We minimize this contribution by excluding data which is significantly modified by ionospheric weather. We determine the quality of the observation using various metrics described in \citet{beardsley_first_2016}, which can be further supplemented by ionospheric data products \citep{jordan_characterization_2017,trott_assessment_2018}.

Not all sources are unresolved. We also can optionally include extended sources by modelling their contribution as a series of unresolved point sources \citep{carroll_thesis}. This can also be done for creating models of diffuse synchrotron emission \citep{beardsley_thesis}. However, the diffuse emission is significantly polarised \citep{lenc_low-frequency_2016} and difficult to include.

With these assumptions, we now have a reliable sky model that we can use to generate model visibilities. For each point source in our catalogue, we perform a discrete Fourier transform using the RA/Dec floating-point location and the Stokes I brightness. This results in a discretised, model $uv$-plane for each source \textit{without} instrumental effects; typically at $\frac{1}{2}$\,$\lambda$ resolution. All $uv$-planes from all sources are summed to create a model $uv$-plane of the sky, and this process is repeated for each observation to minimize $w$-terms associated with the instantaneous measurement plane (see \S\ref{sec:integration} for more discussion).

Once a model $uv$-plane is created with all source contributions, we simulate what the instrument actually measures. The hyperresolved $uv$-beam from \S\ref{subsec:beam} is the sensitivity of the cross-correlation of two elements. We calculate the $uv$-locations of each cross-correlated visibility, and multiply the model $uv$-plane with the $uv$-beam sampling function. The sum of the sensitivity multiplied by the model at the sampled points is the estimated measured value for that cross-correlated visibility.

These visibilities represent our best estimate of what the instrument should have measured, disregarding any source position offsets $\mathbfss{O}$, polarised Faraday rotation $\mathbfss{F}$, and diffuse emission. Any deviations from these model visibilities (whether instrumental or not) will manifest as errors in the comparison between the data and model during calibration.

Our model visibilities can be represented in the signal path framework as
\begin{equation}
    \textbf{m} \sim \mathbfss{S} \   \textbf{I}^{\textrm{true}},
    \label{eq:gridded}
\end{equation}
where $\mathbfss{m}$ are the estimated model visibilities, $\textbf{I}^{\textrm{true}}$ is the true sky, and $\mathbfss{S}$ are contributions from the plane of the measurement. We have not included any modifications in the signal path from before the instrument, $\mathbfss{B}$, and have instead chosen to use an avoidance technique for affected data.

\subsection{Calibration}
\label{subsec:calibration}

We are left with one type of modification to the signal that has not been accounted for by the model visibilities or by avoidance: $\mathbfss{E}$, the electronic response. This is what we classify as our calibration.

At this point in the analysis, we have a measurement equation that looks like
\begin{equation}
\textbf{v}^{\textrm{meas}} \sim \mathbfss{E} \, \mathbfss{m} + \textrm{noise}.
\label{eq:model_meas}
\end{equation}
We assume the electronic response $\mathbfss{E}$ varies slowly with time, and thus does not change significantly over an observation (\textit{e.g.} 2\,minutes for the MWA). Due to our model-based assumptions, we do not have any non-celestial time correlations, antenna correlations, or unknown polarisation correlations. The electronic response is thus simply a time-independent gain $\mathbfss{G}$ per observation which is independent per element and polarisation.

We begin by rewriting Equation~\ref{eq:model_meas} using these assumptions. The measured cross-correlated visibilities are a function of frequency, time, and polarisation. Individual elements are grouped into the sets $A=\{a_0,a_1,a_2,\dots,a_{n}\}$ and $B=\{b_0,b_1,b_2,\dots,b_{n}\}$, where $a$ and $b$ iterate through antenna pairs. A visibility is measured for each polarisation $P=\{p_ap_b, p_aq_b, q_ap_b, q_aq_b\}$, where $p$ and $q$ iterate through the two orthogonal instrumental polarisation components for each element $a$ and $b$.

The resulting relation between the measured visibilities and the model visibilities is
\begin{multline}
    \textbf{v}_{ab,pq} ([f_o,t]) \sim \\ \mathbfss{G}_{a,p}(f_o,f_i) \mathbfss{G}_{b,q}^{*} (f_o,f_i) \textbf{m}_{ab,pq}([f_i,t]) \\+ \textbf{n}_{ab,pq}([f_o,t]),
    \label{fulleq}
\end{multline}
where $\textbf{v}_{ab,pq}([f_o,t])$ are the measured visibilities and $\mathbf{m}_{ab,pq}([f_i,t])$ are the model visibilities. Both are frequency and time vectors $[f,t]$ of the visibilities over all $A$ and $B$ element pairs and over all $P$ polarisation products. $\bm{\mathbfss{G}}_{a,p}(f_o,f_i)$ is a frequency matrix of gains given input frequencies $f_i$ which affect multiple output frequencies $f_o$ for instrumental polarisations $p$ for elements in $A$ (and likewise for $q$ and $B$). Thermal noise $\textbf{n}$ is independent for each visibility. All variables used in this section are summarised in Table\,\ref{tabledef} for reference.

\renewcommand{\arraystretch}{1.1}
\begin{table*}{}
\centering
    \begin{tabular}{c c c}
            Type & Variables & Definition \\
        \hline
            \multicolumn{1}{||c|}{\multirow{4}{*}{\begin{tabular}{@{}c@{}}Observation \\ \& Scale \\ Parameters \end{tabular}}} &
            \multicolumn{1}{c|}{$f$} &
            \multicolumn{1}{l||}{\begin{tabular}{@{}l@{}}Measured frequencies of the observation. \end{tabular}}   \\
        \cline{2-3}
            \multicolumn{1}{||c|}{} &
            \multicolumn{1}{c|}{$p,q$} &
            \multicolumn{1}{l||}{\begin{tabular}{@{}l@{}}Orthogonal instrumental polarisations of the elements in the array.\end{tabular}}  \\
        \cline{2-3}
            \multicolumn{1}{||c|}{} &
            \multicolumn{1}{c|}{$a,b$} &
            \multicolumn{1}{l||}{\begin{tabular}{@{}l@{}}Elements in the array.\end{tabular}}  \\
        \cline{2-3}
            \multicolumn{1}{||c|}{} &
            \multicolumn{1}{c|}{$t$} &
            \multicolumn{1}{l||}{\begin{tabular}{@{}l@{}}Time steps within an observation.\end{tabular}} \\
        \hline
        \hline
            \multicolumn{1}{||c|}{\multirow{10}{*}{Sets}} &
            \multicolumn{1}{c|}{$A=\{a_0,a_1,a_2,...a_{n}\}$} &
            \multicolumn{1}{l||}{\multirow{2}{*}{\begin{tabular}{@{}l@{}}All elements of the array. $A$ and $B$ can be iterated separately to form\\ cross-correlated element pairs.\end{tabular}}}   \\
        \cline{2-2}
            \multicolumn{1}{||c|}{} &
            \multicolumn{1}{c|}{$B=\{b_0,b_1,b_2,...b_{n}\}$} &
            \multicolumn{1}{c||}{}  \\
        \cline{2-3}
            \multicolumn{1}{||c|}{} &
            \multicolumn{1}{c|}{\begin{tabular}{@{}c@{}}$P= \{p_ap_b,...q_aq_b\}$\end{tabular}} &
            \multicolumn{1}{l||}{\begin{tabular}{@{}l@{}}Polarisations in the coherency domain between two elements.\end{tabular}} \\
        \cline{2-3}
            \multicolumn{1}{||c|}{} &
            \multicolumn{1}{c|}{\begin{tabular}{@{}c@{}}$C=\{c_n,...c_0,$\\$\phi_n,...\phi_0\}$\end{tabular}} &
            \multicolumn{1}{l||}{\begin{tabular}{@{}l@{}}Coefficients of a low-order amplitude and phase polynomial across the\\frequency band.\end{tabular}}  \\
        \cline{2-3}
            \multicolumn{1}{||c|}{} &
            \multicolumn{1}{c|}{\begin{tabular}{@{}c@{}}$L=\{l_{0},l_{1},l_{2},...l_{n}\}$\end{tabular}} &
            \multicolumn{1}{l||}{\begin{tabular}{@{}l@{}}Sets of elements associated with cable lengths and types.\end{tabular}} \\
        \cline{2-3}
            \multicolumn{1}{||c|}{} &
            \multicolumn{1}{c|}{
            $D=\{c,\tau,\phi\}$} &
            \multicolumn{1}{l||}{\begin{tabular}{@{}l@{}}Amplitude, mode, and phase of a cable reflection fit across the frequen-\\ cy band.\end{tabular}} \\
        \cline{2-3}
            \multicolumn{1}{||c|}{} &
            \multicolumn{1}{c|}{\begin{tabular}{@{}c@{}}$T=\{\rho_{0},\rho_{1},\rho_2,...\rho_{n}\}$\end{tabular}} &
            \multicolumn{1}{l||}{\begin{tabular}{@{}l@{}}Observation timing sets of physical time separations, such as pointings.\end{tabular}}  \\
        \cline{2-3}
            \multicolumn{1}{||c|}{} &
            \multicolumn{1}{c|}{$[f,t]$} &
            \multicolumn{1}{l||}{\begin{tabular}{@{}l@{}} A combined set of all frequencies and times.\end{tabular}}  \\
        \hline
        \hline
            \multicolumn{1}{||c|}{\multirow{18}{*}{\begin{tabular}{@{}c@{}}Groups, \\ Matrices, \\ \& Vectors\end{tabular}}} &
            \multicolumn{1}{c|}{$\bm{\alpha}_{L}$} &
            \multicolumn{1}{l||}{\begin{tabular}{@{}l@{}}An element grouping, where parameters are per element group rather\\ than per element.\end{tabular}}\\
        \cline{2-3}
            \multicolumn{1}{||c|}{} &
            \multicolumn{1}{c|}{$\bm{\theta}_T$} &
            \multicolumn{1}{l||}{\begin{tabular}{@{}l@{}}A grouping of observation times, where parameters are per timing\\ group rather than per observation time. \end{tabular}}\\
        \cline{2-3}
            \multicolumn{1}{||c|}{} &
            \multicolumn{1}{c|}{$\mathbfss{G}_{a,p}(f_o,f_i)$} &
            \multicolumn{1}{l||}{\begin{tabular}{@{}l@{}}The full gain matrix for each element in group $A$ per $p$ where input\\ and output frequencies are correlated. \end{tabular}}\\
        \cline{2-3}
            \multicolumn{1}{||c|}{} &
            \multicolumn{1}{c|}{$\textbf{g}_{a,p}(f)$} &
            \multicolumn{1}{l||}{\begin{tabular}{@{}l@{}} A vectorised approximation of the gains $\mathbfss{G}$ for each element in the\\ group $A$ per $P$ over frequency.\end{tabular}}\\
        \cline{2-3}
            \multicolumn{1}{||c|}{} &
            \multicolumn{1}{c|}{$\textbf{m}_{ab,pq}([f,t])$} &
            \multicolumn{1}{l||}{\begin{tabular}{@{}l@{}} A vector of the simulated model visibilities from a model sky with \\frequency-dependent beam effects for each element pair $ab$ and \\polarisation product $pq$ over the set $[f,t]$.\end{tabular}}\\
        \cline{2-3}
            \multicolumn{1}{||c|}{} &
            \multicolumn{1}{c|}{$\textbf{n}_{ab,pq}([f,t])$} &
            \multicolumn{1}{l||}{\begin{tabular}{@{}l@{}} A vector of the thermal noise for each element pair $ab$ and polarisation\\ product $pq$ over the set $[f,t]$.\end{tabular}}\\
        \cline{2-3}
            \multicolumn{1}{||c|}{} &
            \multicolumn{1}{c|}{$\textbf{v}_{ab,pq}([f,t])$} &
            \multicolumn{1}{l||}{\begin{tabular}{@{}l@{}} A vector of the uncalibrated data visibilities for each element pair $ab$\\ and polarisation product $pq$ over the set $[f,t]$.\end{tabular}}\\
        \cline{2-3}
            \multicolumn{1}{||c|}{} &
            \multicolumn{1}{c|}{$\textbf{a}_{a,p}(f)$} &
            \multicolumn{1}{l||}{\begin{tabular}{@{}l@{}} The calculated auto-gain for each element in the group $A$ per $p$ over \\ frequency.\end{tabular}}\\
        \cline{2-3}
            \multicolumn{1}{||c|}{} &
            \multicolumn{1}{c|}{$\bm{\eta}_{a,p}(f)$} &
            \multicolumn{1}{l||}{\begin{tabular}{@{}l@{}}Scaling relation of the discrepancy between the measured sky and the \\ calibration model in the cross-correlations.\end{tabular}} \\
        \cline{2-3}
            \multicolumn{1}{||c|}{} &
            \multicolumn{1}{c|}{$\hat{\textbf{a}}_{a,p}(f)$} &
            \multicolumn{1}{l||}{\begin{tabular}{@{}l@{}} The calculated auto-gain for each element in the group $A$ per $p$ over \\ frequency which has been scaled to match the  cross-correlations.\end{tabular}}\\
        \hline
        \hline
            \multicolumn{1}{||c|}{\multirow{3}{*}{Functions}} &
            \multicolumn{1}{c|}{$\mathcal{P}(\ )$} &
            \multicolumn{1}{l||}{\begin{tabular}{@{}l@{}} A polynomial fit as a function of frequency of the input. \end{tabular}}\\
        \cline{2-3}
            \multicolumn{1}{||c|}{} &
            \multicolumn{1}{c|}{$\mathcal{R\langle \ \rangle}$} &
            \multicolumn{1}{l||}{\begin{tabular}{@{}l@{}} A resistant mean of the input vector over element set $L$ (and option-\\ally a time group $\theta_T$).  Outliers beyond $2\sigma$ are excluded in the average. \end{tabular}}\\
        \hline
        \hline
            \multicolumn{1}{||c|}{\multirow{7}{*}{\begin{tabular}{@{}c@{}}Discrete\\ Fourier\\ Transforms \end{tabular}}} &
            \multicolumn{1}{c|}{$k$} &
            \multicolumn{1}{l||}{\begin{tabular}{@{}l@{}}  The Nyquist frequency index of the Fourier dual of frequency.\end{tabular}}\\
        \cline{2-3}
            \multicolumn{1}{||c|}{} &
            \multicolumn{1}{c|}{$\kappa$} &
            \multicolumn{1}{l||}{\begin{tabular}{@{}l@{}} The hyperfine sub-Nyquist frequency index of the Fourier dual of\\ frequency, with resolution at $1 / 20^{\textrm{th}}$ of $k$. \end{tabular}}\\
        \cline{2-3}
            \multicolumn{1}{||c|}{} &
            \multicolumn{1}{c|}{$n$} &
            \multicolumn{1}{l||}{\begin{tabular}{@{}l@{}} The index of the frequency. \end{tabular}}\\
        \cline{2-3}
            \multicolumn{1}{||c|}{} &
            \multicolumn{1}{c|}{$N$} &
            \multicolumn{1}{l||}{\begin{tabular}{@{}l@{}} The total number of frequency channels. \end{tabular}}\\
        \cline{2-3}
            \multicolumn{1}{||c|}{} &
            \multicolumn{1}{c|}{$\tau_\kappa$} &
            \multicolumn{1}{l||}{\begin{tabular}{@{}l@{}} The Fourier dual of frequency: a timing delay in the detection of the \\waveform between one element and another. The $\kappa$ index indicates it \\runs over the hyperfine index. \end{tabular}}\\
        \hline

     \end{tabular}
    \caption{Definitions of the variables used for calibration, organised by type.}
    \label{tabledef}

\end{table*}

Our notation has been specifically chosen. Naturally discrete variables (element pairs and polarisation products) are described in the subscripts. Naturally continuous variables (frequency and time) are function arguments. We group frequency and time into a set $[f,t]$ in the visibilities to create vectors. Since frequency and time are independent, this notation is more compact. In contrast, the gain matrices are not independent in frequency. A full matrix must be used to accurately capture frequency correlation due to $\mathbfss{A}$, which includes aliasing from common electronics like polyphase filter banks or bandpasses. 

To reduce Equation~\ref{fulleq} significantly, we make the assumption that the frequencies are independent in $\mathbfss{G}$. This forces the frequency correlation contribution $\mathbfss{A} = \mathbb{I}$, giving
\begin{equation}
    \mathbfss{G}_{a,p}(f_o,f_i) \sim \diag{\textbf{g}_{a,p}(f)}.
    \label{Gest}
\end{equation}
The instrumental gains $\textbf{g}$ are now an independent vector of frequencies for elements in $A$ per instrumental polarisations $p$ (and likewise for $q$ and $B$). We flag frequency channels which are most affected by aliasing to make this assumption viable. The effect of this flagging is dependent on the instrument; for the MWA, flagging every 1.28\,MHz to remove aliasing from the polyphase filter banks introduces harmonic contamination in the power spectrum. Thus, enforcing this assumption usually has consequences in the power spectrum space.

We can now fully vectorise the variables in Equation~\ref{fulleq}:
\begin{multline}
    \textbf{v}_{ab,pq} ([f,t]) \sim \\  \diag{\textbf{g}_{a,p}(f)} \diag{\textbf{g}_{b,q}^{*} (f)} \textbf{m}_{ab,pq}([f,t])\\ + \textbf{n}_{ab,pq}([f,t]).
    \label{vector}
\end{multline}
The gains calculated from solving Equation~\ref{vector} will encode differences between the model visibilities and the true visibilities. This is a systematic; the gains will be contaminated with a non-instrumental contribution. We can reduce this effect with long-baseline arrays by excluding short baselines from calibration since we lack a diffuse-emission model. However, this can potentially introduce systematic biases at low $\lambda$ \citep{patil_systematic_2016}. For the remainder of \S\ref{subsec:calibration}, we describe our attempts to remove the systematics encoded within the gains due to an imperfect model.


\subsubsection{Per Frequency Solutions}
\label{least-squares}

Equation~\ref{vector} can be used to solve for the instrumental gains for all frequencies and polarisations independently. This allows the use of Alternating Direction Implicit (ADI) methods for fast and efficient solving of $\mathcal{O}(N^2)$ \citep{mitchell_real-time_2008,salvini_fast_2014}. Due to this independence, parallelisation can also be applied. Noise is ignored during the ADI for simplicity; the least-squares framework is the maximum likelihood estimate for a Gaussian distribution. However, if any noise has a non-Gaussian distribution, it will affect the instrumental gains.

We begin solving Equation~\ref{vector} by estimating an initial solution for $\textbf{g}_{b,q}^{*}(f)$ to force the gains into a region with a local minimum. Reliable choices are the average gain expected across all elements or scaled auto-correlations.

With an input for $\textbf{g}_{b,q}^{*}(f)$, Equation~\ref{vector} can then become a linear least-squares problem.
\begin{multline}
    \bm{\chi}^{2}_{a,p}([f,t]) = \sum_{b} \Big| \textbf{v}_{ab,pq} ([f,t]) - \\ \diag{\textbf{g}_{a,p}(f)}~\diag{\textbf{g}_{b,q}^{*}(f)}~\textbf{m}_{ab,pq}([f,t]) \Big|^2,
    \label{sum}
\end{multline}
where $\textbf{g}_{a,p}(f)$ is found given a minimisation of $\bm{\chi}^{2}_{a,p}([f,t])$ for each element $a$ and instrumental polarisation $p$. All time steps are used to find the temporally constant gains over the observation. For computation efficiency, we have also assumed $P = \{p_ap_b,q_aq_b\}$ (e.g $XX$ and $YY$ in linear polarisation) since these contributions are most significant. Optionally, a full polarisation treatment can be used given polarised calibration sources.

The current estimation of $\textbf{g}_{b,q}^{*}(f)$ is then updated with knowledge from $\textbf{g}_{a,p}(f)$ by adding together the current and new estimation and dividing by 2. By updating in partial steps, a smooth convergence is ensured. The linear least-squares process is then repeated with an updated $\textbf{g}_{b,q}^{*}(f)$ until convergence is reached\footnote{We have found that allowing the first 10 iterations to only update the phase of $\textbf{g}_{b,q}^{*}(f)$ helps to converge faster.}.

\subsubsection{Bandpass}
\label{bandpass}

The resulting gains from the least-squares iteration process are fully independent in frequency, element, observation, and polarisation. This is not a completely accurate representation of the gains. It was necessary to make this assumption for the efficient solving technique in \S\ref{least-squares}, but we can incorporate our prior knowledge of the nature of the instrument and its spectral structure \emph{ex post facto}.

\begin{figure}
\centering
	\includegraphics[width = \columnwidth]{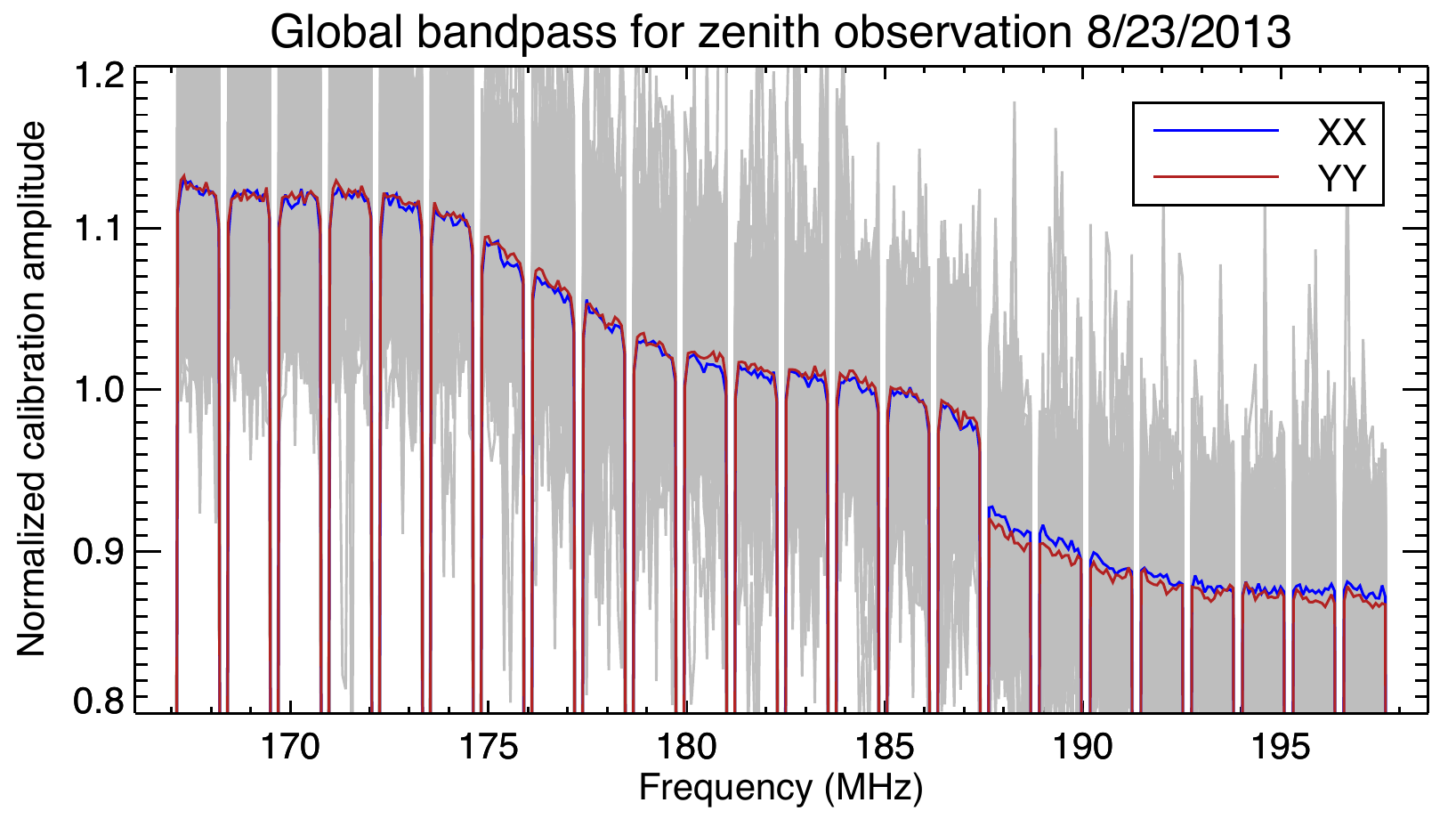}
	\caption[Global bandpass for a zenith observation]{The MWA global bandpass for the zenith observation of August 23, 2013 for polarisations $pp = XX$ (blue) and $qq = YY$ (red). All the per-frequency antenna solutions used in the global bandpass average are shown in the background (grey). This historical approach greatly decreased expected noise on the solutions.}
	\label{fig:global_bp}
\end{figure}

For example, we did not account for noise contributions during the per-frequency ADI fit; this adds spurious deviations from the gain's true value with mean of zero. Historically, we accounted for these effects by creating a global bandpass,
\begin{equation}
    \left|\textbf{g}_{p}(f; \bm{\alpha})\right|  = \left\langle \left|\textbf{g}_{a,p}(f)\right| \right\rangle_{\bm{\alpha}},
    \label{eq:global_bp}
\end{equation}
where the normalised amplitude average is taken over all elements $\bm{\alpha}$ as a function of frequency to create a global bandpass $\left|\textbf{g}_{p}(f; \bm{\alpha})\right|$ independent of elements. Figure~\ref{fig:global_bp} shows an example of the global bandpass alongside the noisy per-frequency inputs for the MWA.

This methodology drastically reduces noise contributions to the bandpass when there are many elements, and it reduces spectral structure contributions due to imperfections in the model \citep{barry_calibration_2016}. However, this averaging implicitly assumes that all elements are identical. We must use other schemes to capture more instrumental parameters while maintaining spectral smoothness in the bandpass. 

The level of accuracy required in the calibration to feasibly detect the EoR is 1 part in $10^5$ as a function of frequency \citep{barry_calibration_2016}. As seen in Figure~\ref{fig:global_bp}, instruments can be very spectrally complicated. Therefore, more sophisticated calibration procedures must be used. 

Usually, these schemes are instrument-specific. For example, in the MWA, sets of elements experience the same attenuation as a function of frequency due to cable types, cable lengths, and whitening filters. We group these elements into different cable length sets $L=\{l_{0},l_{1},l_{2},...l_{n}\}$. We get
\begin{equation}
    \left|\textbf{g}_{l,p}(f; \bm{\alpha}_{l})\right| = \mathcal{R}\left\langle   \left|\textbf{g}_{a\in L,p}(f)\right|,2\sigma  \right\rangle,
\end{equation}
where $\mathcal{R}$ is the resistant mean function\footnote{The resistant mean function calculates the distribution of the amplitudes of a similar element set $\bm{\alpha}_l$ for each frequency and polarisation and then calculates the mean of that distribution after Gaussian $2\sigma$ outliers have been excluded.} calculated over each element set $\bm{\alpha}_L$ for each polarisation and frequency. We choose the resistant mean because outlier contributions are more reliably reduced than median calculations.  The variable change from $a$ to $\bm{\alpha}$ indicates one parameter per group of elements.

If more observations are available, we follow a similar averaging process over time. If the instrument is stable in time, a normalised bandpass per antenna should be nearly identical from one time to the next excluding noise contributions and potential Van Vleck quantisation corrections \citep{vleck_spectrum_1966}. Gains from different LSTs will have different spectral structure from unmodelled sources, and thus an average will remove even more of this effect.

We create a time set  $T=\{\rho_{0},\rho_{1},\rho_2,...\rho_{n}\}$ where times are grouped by physical time separations based on the instrument. For example, we group the MWA observations by pointings\footnote{A pointing defines a group of observations with the same electronic delay. As the sky rotates throughout the night, different electronic delays are used to roughly point the instrument to the same location in the sky.} because they sample different beam errors, and thus averaging between pointings would remove this instrumental feature. Whenever possible, we use
\begin{equation}
    \left|\textbf{g}_{l,p}(f; \bm{\alpha}_{l}, \bm{\theta}_\rho)\right| = \mathcal{R}\left\langle \left|\textbf{g}_{a\in L,p}(f)\right|,t \in T,2\sigma \right\rangle,
\end{equation}
where $\bm{\theta}_\rho$ runs over observations within physical time separations in the set $T$ and over as many days as applicable. The variable change from $t$ to $\bm{\theta}$ indicates one parameter per group of times.

\subsubsection{Low-order polynomials}
\label{polys}

An overall amplitude due to temperature dependence $\mathbfss{T}$ in the amplifiers must still be accounted for within the gains. These differ from day to day and element to element, therefore they cannot be included in the average bandpass. For well-behaved amplifiers, this varies slightly as a function of frequency and is easily characterised with a low-order polynomial for each element and observation.

In addition to fitting polynomials to the amplitude as a function of frequency, we must also account for the phase. There is an inherent per-frequency degeneracy that cannot be accounted for by sky-based calibration. Therefore, we reference the phase to a specific element to remove this degeneracy, which makes the collective phases smooth as a function of frequency for well-behaved instruments. We have found that using a per-element polynomial fit as the calibration phase solution has been a reliable estimate.

For the amplitude, we fit
\begin{equation}
    c_{n}f^n+...+c_{1}f+c_{0} = \mathcal{P}\left(  \frac{|\textbf{g}_{a,p}(f)|} {|\textbf{g}_{p}(f; \bm{\alpha})|} \right),
    \label{eq:amp_poly}
\end{equation}
where the bandpass contribution, $|\textbf{g}_{p}(f; \bm{\alpha})|$, is removed before the fit and $c_{n}$ are the resulting coefficients. Any bandpass contribution can be used; $|\textbf{g}_{p}(f; \bm{\alpha})|$ is just an example. For the phase, we fit
\begin{equation}
    \phi_{n}f^n+...+\phi_{1}f+\phi_{0} = \mathcal{P}\left( \arg{\textbf{g}_{a,p}(f)} \right),
    \label{eq:phase_poly}
\end{equation}
where the polynomial fit is done over the phase of the residual and $\phi_{n}$ are the resulting coefficients. Due to phase jumps between $-\pi$ and $\pi$, special care is taken to ensure the function is continuous across the $\pi$ boundary\footnote{We ``unwrap'' the phase by creating a new continuous plane from the Riemann sheets. We then solve, and ``rewrap.'' If the phase varies quickly, there can be ambiguity in which Riemann sheet to place the phase, but this is not an issue with most instruments.}. We can create a set of these coefficients, $C=\{c_n,...c_1, c_0, \phi_n,...\phi_1, \phi_0\}$, for easy reference.

In all cases, we choose the lowest-order polynomials possible for these fits on the amplitude and phase. For example, the MWA is described by a $2^\textrm{nd}$-order polynomial fit in amplitude and a linear fit in phase. As a general rule, we do not fit polynomials which have modes present in the EoR window.

\subsubsection{Cable reflections}
\label{cableref}

Reflections due to a mismatched impedance must also be accounted for within the gains (contribution $\mathbfss{K}$ in the signal path framework). Even though instrumental hardware is designed to meet engineering specifications, residual reflection signals are still orders of magnitude above the EoR and are different for each element signal path. Averaging the gains across elements and time in \S\ref{bandpass} artificially erased cable reflections from the solutions, thus we must specifically incorporate them. Currently, we only fit the cable reflection for elements with cable lengths that can contaminate prime locations within the EoR window in power spectrum space \citep{beardsley_first_2016,ewall-wice_first_2016}.

We find the theoretical location of the mode using the nominal cable length and the specified light travel time of the cable. We then perform a hyperfine discrete Fourier transform of the gain around the theoretical mode
\begin{multline}
    \textbf{g}_{a,p}(\tau_\kappa) = \sum_{m=0}^{M-1} \Bigg(\frac{\textbf{g}_{a,p}(f_m)} {|\textbf{g}_{p}(f; \bm{\alpha})| } \\ -  (c_{n}f_m^n+...+c_{0}) \ e^{i\left(\phi_{n}f_m^n+...+\phi_{0}\right)}\Bigg) \ e^{-2\pi i \kappa \frac{m}{M}},
\end{multline}
where $\tau_\kappa$ is the delay, $[m,M] \in\mathbb{Z}$, and $\kappa$ is the hyperfine index component. Typically, we set the range of $\kappa$ to be $[k_{\tau_{o}}-\frac{1}{20}k, k_{\tau_{o}}+\frac{1}{20}k]$, where $k_{\tau_{o}}$ is the index of the theoretical mode and $k$ is the index in the range of $[0,M-1]$ \citep{beardsley_thesis}. Again, any bandpass contribution can be used; $|\textbf{g}_{p}(f; \bm{\alpha})|$ is just an example.

The maximum $|\textbf{g}_{a,p}(\tau_\kappa)|$ around $k_{\tau_{o}}$ is chosen as the experimental cable reflection. The associated amplitude $c$, phase $\phi$, and mode $\tau$ are then calculated to generate the experimental cable reflection contribution $c e^{-2 \pi i \tau f + i \phi}$ to the gain for each observation. Optionally, these coefficients can be averaged over sets of observations and times, depending on the instrument. We can create a set of these coefficients, $D=\{c, \tau, \phi\}$, for easy reference.

\subsubsection{Auto-correlation bandpass}
\label{subsubsec:autos}

The fine-frequency bandpass is a huge potential source of error in the power spectrum \citep{barry_calibration_2016}. As such, we have tried averaging along various axes to reduce implicit assumptions while maintaining spectral smoothness in \S\ref{bandpass}. However, there always remains an assumption of stability along any axis that we average over, and this may not be valid enough for the level of required spectral accuracy.

We can instead use the auto-correlations to bypass these assumptions. An auto-correlation should only contain information about the total power on the sky, instrumental effects, and noise. No information about structure on the sky is encoded, so spectral effects from unmodelled sources do not contribute. However, there are two main issues with using auto-correlations as the bandpass: improper scaling and correlated noise.

\begin{itemize}
  \item The scale of the auto-correlations are dominated by the largest modes on the sky. Currently, we do not have reliable models of these largest modes. Either a large-scale calibration model must be obtained, or the scaling must be forced to match the cross-correlations.
  \item Thermal noise is correlated in an auto-correlation, and will contribute directly to the solutions. Truncation effects during digitisation and channelisation can artificially correlate visibilities. This will result in bit noise, which will also contribute to the auto-correlations.
\end{itemize}

Whether or not auto-correlations can effectively be used for the bandpass depends on the importance of their errors in power spectrum space and our ability to mitigate these errors.
We solve for the auto-correlation gains via the relation,
\begin{equation}
    \textbf{a}_{a,p}(f) = \sqrt{\frac{ \left\langle  \textbf{v}_{aa,pp} ([f,t]) \right\rangle_t}{\left\langle  \textbf{m}_{aa,pp} ([f,t]) \right\rangle_t}},
\end{equation}
where $\textbf{a}_{a,p}(f)$ is the auto-gain for element $a$ and polarisation $p$ as a function of frequency.

To correct for the scaling error and for the correlated thermal noise floor, we define a cross-correlation scaling relation,
\begin{equation}
    \bm{\eta}_{a,p}(f) = \frac{\mathcal{P}\left( |\textbf{g}_{a,p}(f; \bm{\alpha}_{l},\bm{\theta}_\rho,[C],[D])| \right)}{\mathcal{P}\left( \textbf{a}_{a,p}(f) \right)},
\end{equation}
where $\mathcal{P}$ is a polynomial fit to the gains, usually limited to just a linear fit. This captures the discrepancy between the measured sky and the calibration model in the space of one cross-correlation gain. We then rescale the auto-gains to match the cross-correlations,
\begin{equation}
    \hat{\textbf{a}}_{a,p}(f)  =  \bm{\eta}_{a,p}(f) \ \textbf{a}_{a,p}(f),
\end{equation}
where $\hat{\textbf{a}}_{a,p}(f)$ is the scaled auto-gain for element $a$ and polarisation $p$ as a function of frequency.

We have reduced scaling errors and correlated thermal noise on the auto-gain solutions. However, we have not corrected for bit noise, or a bias caused by bit truncation at any point along the signal path. These errors are instrument-specific and will persist into power spectrum space. If the bits in the digital system are not artificially correlated due to bit truncation, then the auto-correlation can yield a better fine-frequency bandpass.

\subsubsection{Final calibration solutions}

Our final calibration solution using only cross-correlations is
\begin{multline}
    \textbf{g}_{a,p}(f; \bm{\alpha}_{l},\bm{\theta}_\rho,[C],[D]) =  \underbrace{\left|\textbf{g}_{l,p}(f; \bm{\alpha}_{l},\bm{\theta}_\rho)\right|}_\textbf{bandpass \S\ref{bandpass}} \\ \underbrace{\big((c_{n}f^n+...+c_{0}) e^{i\left(\phi_{n}f^n+...+\phi_{0}\right)}}_\textbf{low-order polynomials \S\ref{polys}} +  \underbrace{ce^{-2 \pi i \tau f + i \phi}\big)}_\textbf{reflection \S\ref{cableref}},
    \label{150}
\end{multline}
for elements with fitted cable reflections, and
\begin{multline}
    \textbf{g}_{a,p}(f; \bm{\alpha}_{l},\bm{\theta}_\rho,[C]) = \\
    \underbrace{\left|\textbf{g}_{l,p}(f; \bm{\alpha}_{l},\bm{\theta}_\rho)\right|}_\textbf{bandpass \S\ref{bandpass}} \ \underbrace{(c_{n}f^n+...+c_{0}) e^{i\left(\phi_{n}f^n+...+\phi_{0}\right)}}_\textbf{low-order polynomials \S\ref{polys}},
    \label{eq:cross-no150}
\end{multline}
for all other elements. The bandpass amplitude solution is generated over a set of elements with the same cable/attenuation properties ($\bm{\alpha}_{l}$) and includes many observations within a timing set ($\bm{\theta}_\rho$) covering many days. The same bandpass is applied to all elements of the appropriate type and all observing times from the respective time set. In contrast, the polynomials and the cable reflection are fit independently for each observation and per element. We divide the data visibilities by the applicable form of $\textbf{g}_a\textbf{g}_b^*$ to form our final, calibrated visibilities.

Alternatively, our final calibration solution harnessing the auto-correlations is
\begin{multline}
    \textbf{g}_{a,p}(f; \bm{\alpha}_{l},\bm{\theta}_\rho,[C],[D], \hat{\textbf{a}}) = \\ \underbrace{\hat{\textbf{a}}_{a,p}(f)}_\textbf{auto-gain \S\ref{subsubsec:autos}} \underbrace{e^{i\arg{\textbf{g}_{a,p}(f; \bm{\alpha}_{l},\bm{\theta}_\rho,[C],[D])}}}_\textbf{phase from Eq. \ref{150}} ,
    \label{eq:autos}
\end{multline}
where the amplitude is described by the scaled auto-gains and the phase is described by the cross-gains.

For EoR science, we want to reduce spectral structure as much as possible but still capture instrumental parameters. Our auto-gain calibration solution performs the best \textit{in the most sensitive, foreground-free regions}. Therefore, we currently use Equation~\ref{eq:autos} in creating EoR upper limits with the MWA. This is an active area of research within the EoR community; reaching the level of accuracy required to detect the EoR is ongoing.

\subsection{Imaging}
\label{subsec:images}

The final stage of FHD transforms calibrated data visibilities into a space where they can be combined across observations. To begin this process, we perform an operation called gridding on the visibilities.

We take each complex visibility value and multiply it by the corresponding $uv$-beam in the coherency domain calculated in \S\ref{subsec:beam} through a process called Optimal Map-Making. In essence, this takes visibility values integrated by the instrument and estimates their original spatial $uv$-response given our knowledge of the beam. These are approximations of the instrument power response for each baseline to a set of regular gridding points (e.g. \citet{myers_fast_2003,bhatnagar_correcting_2008,morales_software_2009, sullivan_fast_2012, dillon_overcoming_2014, shaw_all-sky_2014, zheng_brute-force_2017}).

We separately perform gridding to individual $uv$-planes for calibrated data visibilities, model visibilities, and residual visibilities generated from their difference. By gridding each separate data product, we can make a variety of diagnostic images. For example, images generated from residual visibilities help to ascertain the level of foreground removal in image space, and thus are important for quality assurance.

We also grid visibilities of value 1 with the beam gridding kernel to create natural $uv$-space weights. This generates a sampling map which describes how much of a measurement went into each pixel. In addition, we separately grid with the beam-squared kernel to create a variance map. The variance map relates to the uncertainty for each $uv$-pixel, which will be a vital component for end-to-end error propagation in {\eppsilon}.

We then have three types of $uv$-plane products: the sampling map, the variance map, and the data $uv$-planes. From these, we can create weighted-data $uv$-planes using the sampling map and data  $uv$-planes. All these products are transformed via 2D FFTs to image space in slant orthographic projection\footnote{Slant orthographic projection: a flat projection of the sky that is slanted to be parallel with the measurement plane.}.

\begin{figure*}
\centering
	\includegraphics[width = .7\textwidth]{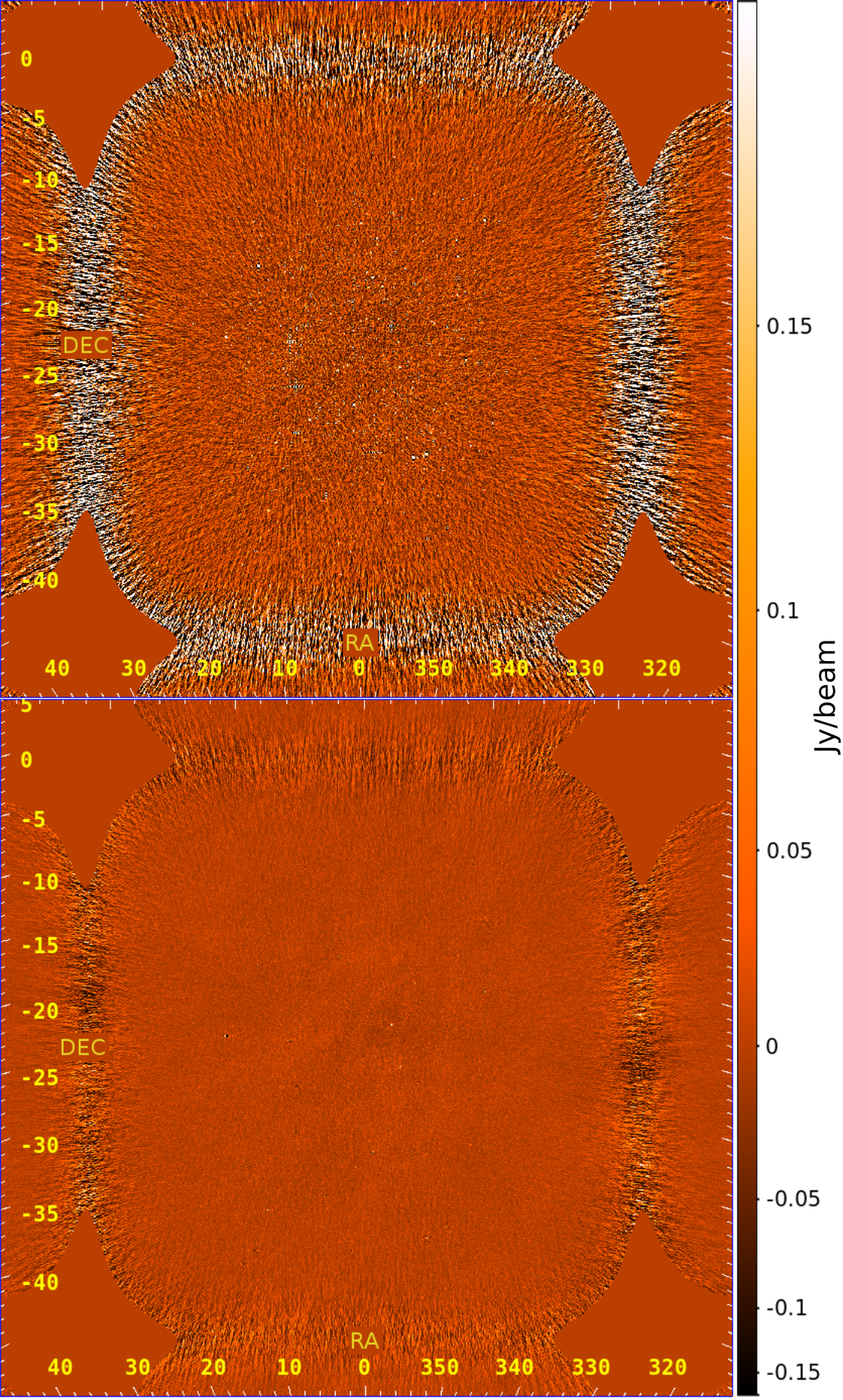}
	\caption[Calibrated data and residual images for a zenith observation]{An example of images output from FHD: calibrated data (top) and residual (bottom) Stokes I images from the MWA for the zenith observation of August 23, 2013. There is significant reduction of sources and point spread functions in the residual. However, the diffuse synchrotron emission can be seen in the residual because it was not in the subtraction model.}
	\label{fig:snapshot_images}
\end{figure*}

At this point, the various images are made for two different purposes: 1) the sampling map, variance map, and data image planes are for power spectrum packages and 2) the weighted-data image plane is for diagnostic images per observation. The result of calibrated data and residual snapshot images for a zenith observation with Phase I of the MWA is shown in Figure~\ref{fig:snapshot_images}.

Our slant orthographic images are in a basis that changes with LST. Therefore, we perform a bilinear interpolation to HEALPix pixel centers\footnote{HEALPix: the Hierarchical Equal Area isoLatitude Pixelization of a sphere.} \citep{gorski_healpix:_2005}, which are the same for all LSTs. We interpolate the calibrated data, model, sampling map, and variance map to HEALPix pixel centers \textit{separately} for use in {\eppsilon}. If we want to combine multiple observations, we will need to do a weighted average. Therefore, we keep the numerator (data) separate from the denominator (sampling map) for this purpose.

\subsection{Interleaved cubes}
\label{subsec:interleaved_cubes}

The images that will be used for power spectrum analysis are split by interleaved time steps, grouped by even index and odd index. The even--odd distinction is arbitrary; what really matters is that they are interleaved. While this doubles the number of Fourier transforms to perform, it allows for crucial error analysis.

The sky should not vary much over sufficiently small integration intervals. Any significant variation can be attributed to either RFI (which has been accounted for in \S\ref{subsec:flagging}) or thermal noise on the observation. By subtracting the even--odd groupings and enforcing consistent flagging, we should be left with the thermal noise contribution to the observation.

We carry even--odd interleaved cubes throughout the power spectrum analysis and check at various stages against noise calculations. It is a robust way to ensure that error propagation and normalisation has occurred correctly, and it allows us to build more diagnostic data products while still building the cross-power spectrum (see \S\ref{subsec:Various_power_spectrum_products}).

Thus, there are at least eight images per polarisation product output from FHD for use in a power spectrum analysis: the calibrated data, model, sampling map, and variance map for each even--odd set.

\section{Integration}
\label{sec:integration}

In order to reduce noise to reach the signal-to-noise required to detect the EoR, we must now integrate. This requires integration of thousands of observations \citep{beardsley_eor_2013}.

FHD outputs a variety of data cubes that we must integrate together before input into our next package, {\eppsilon} (described in \S\ref{sec:eppsilon}). These RA/Dec frequency slices are numerators and denominators which create meaningful maps in $uv$-space. The various cubes are:
\begin{description}
    \item \textbf{Calibrated data cubes}: HEALPix images for each frequency of the unweighted calibrated data.
    \item \textbf{Model data cubes}: HEALPix images for each frequency of the unweighted generated model. 
    \item \textbf{Sampling map cubes}: HEALPix images for each frequency of the sampling estimate.
    \item \textbf{Variance map cubes}: HEALPix images for each frequency of the variance estimate.
\end{description}
These cube types are split by instrumental polarisation and by an interleaved time sampling set of even and odd time indices (detailed in \S\ref{subsec:interleaved_cubes}). Each even--odd polarisation grouping is added across observations, pixel by pixel, using the HEALPix coordinate system. Averaging \textit{in image space}, rather than in $uv$-space, is a crucial aspect of the analysis. Either approach can theoretically be used, but the computational requirements vary greatly.

The measurement plane is not parallel to the tangent plane of the sky for non-zenith measurements. The instrument's measurement plane appears tilted compared to the wave's propagation direction; this tilt causes a measurement delay. Since each measurement plane will measure different phases of the wave propagation, different $uv$-planes cannot be coadded. This is a classic decoherence problem in Fourier space.

There are two methodologies to account for this decoherence. The first is to project the measurement $uv$-plane to be parallel to the tangent plane of the sky. This is called $w$-projection since it projects $\{u,v,w\}$-space to $\{u,v,w=0\}$-space \citep{cornwell_noncoplanar_2008}. Unfortunately, the projection requires the propagation of a Fresnel pattern for every visibility to reconstruct the wave on the $w=0$ plane. This requires intense computational overhead, but has been successfully implemented by other packages \citep{trott_chips:_2016}.

The second methodology is to integrate in image space. We create an image for each observation by performing a 2D spatial FFT of the $uv$-plane, which results in a slant orthographic projection of the sky. This inherently assumes the array is coplanar; the integration time must be relatively small and altitude variations must be insignificant or absorbed into unique kernel generation (\S\ref{subsec:beam}). For the MWA Phase I, a mean $w$-offset of $0.15\,\lambda$ in the imaged modes is small enough to not be a dominating systematic for current analyses.

We can then easily interpolate from the slant orthographic projection to the HEALPix projection \citep{ord_interferometric_2010}, which is a constant basis. We do not need to propagate waves in image space in order for coherent integration, thus it is computationally efficient in comparison. 

\section{Error Propagated Power Spectrum with Interleaved Observed Noise}
\label{sec:eppsilon}

Error Propagated Power Spectrum with Interleaved Observed Noise, or {\eppsilon}\footnote{\url{https://github.com/EoRImaging/eppsilon}}, is an open source power spectrum analysis package designed to take integrated images and create various types of diagnostic and limit power spectra. It was created as a way to propagate errors into power spectrum space, rather than have \textit{estimated} errors. Other image-based power spectrum analyses exist for radio interferometric data \citep{paciga_gmrt_2011,shaw_all-sky_2014,patil_constraining_2014,shaw_coaxing_2015,dillon_empirical_2015,ewall-wice_first_2016,patil_upper_2017}, however, the end-to-end error propagation of {\eppsilon} and CHIPS \citep{trott_chips:_2016} is uncommon.

The main functions of {\eppsilon} are to transform integrated images into $\{u,v,f\}$-space (\S\ref{subsec:back_to_uv}), calculate observed noise using even--odd interleaving (\S\ref{subsec:mean_and_noise_calculation}), transform frequency to $k$-space (\S\ref{subsec:tTransforming_frequency_to_k-space}), and average $k$-space voxels together for diagnostic and limit power spectra (\S\ref{subsec:Various_power_spectrum_products}).  Detailing these processes will build the groundwork needed to describe 2D power spectra, 1D power spectra, and their respective uncertainty estimates in \S\ref{sec:power_spectrum_diagnostics}.

\subsection{From integrated images to $uv$-space}
\label{subsec:back_to_uv}

The input products of {\eppsilon} are integrated image cubes per frequency. While this space was necessary for integration, the error bars in image space are complicated; every pixel is covariant with every other pixel. Our measurements were inherently taken in $uv$-space, and that is where our uncertainties on the measurement are easiest to propagate.



We then Fourier transform the integrated image back into $uv$-space using a direct Fourier transform between the curved HEALPix sky and the flat, regularly spaced $uv$-plane. Since this happens once per observation integration \textit{set,} rather than once per observation, the slower direct Fourier transform calculation is computationally feasible. The modified gridding kernel  described in Section~\ref{subsec:beam} acts as a window in the HEALPix image. In order to propagate our uncertainties correctly, this must be done during the generation of the kernel.

After transforming into $uv$-space, we calculate the resulting $\{k_x,k_y\}$-values for each pixel. The $uv$-space is related to the wavenumber space by the simple transforms \citep{morales_toward_2004}
\begin{equation}
    k_x = \frac{u2\pi}{D_M(z)} ~~ ~~ ~~ k_y = \frac{v2\pi}{D_M(z)},
\end{equation}
where $D_M(z)$ is the transverse comoving distance dependent on redshift \citep{hogg_distance_1999}.

In addition, we also perform the 3D pixel-by-pixel subtraction of the integrated model from the integrated data to create residual cubes. We could instead perform this after we have transformed $f$ to $k_z$ to save on computation time, but it is helpful to make diagnostic plots of the residual cube in $\{u,v,f\}$-space. By creating diagnostics at every important step in the analysis, we can better understand our systematics.

\subsection{Mean and noise calculation}
\label{subsec:mean_and_noise_calculation}


We now have twenty various $uv$-products: calibrated, model, residual, sampling map, and variance map data, for each polarisation product $pp$ and $qq$, and for each interleaved even--odd time sample set. We have kept the numerators and denominators (data and weights) separate up until this point so that we can perform variance-weighted sums and differences.

First, we weight the calibrated data, model, and residual by the sampling map, thereby upweighting well-measured modes and downweighting poorly measured modes. Second, we weight the variance map by the square of the sampling map to scale our uncertainty estimates with our choice of weighting scheme. We then perform sums and differences using our sampling-map-weighted uncertainty estimates as the weights,
\begin{equation}
    \hat{\mu} = \frac{\frac{x_e}{\sigma_e^2} + \frac{x_o}{\sigma_o^2} }{\frac{1}{\sigma_e^2}+\frac{1}{\sigma_o^2}} ~~ ~~ ~~ \hat{n} = \frac{\frac{x_e}{\sigma_e^2}- \frac{x_o}{\sigma_o^2} }{\frac{1}{\sigma_e^2}+\frac{1}{\sigma_o^2}},
    \label{eq:ml}
\end{equation}
where $\hat{\mu}$ is the mean, $\hat{n}$ is the noise, $\{e,o\}$ are the interleaved even--odd sets, $x$ is the sampling-map-weighted data (calibrated data, model, or residual) for a given even--odd set, and $\sigma^2$ is the sampling-map-weighted variance map for a given even--odd set. The calculated mean and noise are the maximum likelihood estimates for a weighted Gaussian probability distribution, hence using the variables $\hat{\mu}$ and $\hat{n}$ instead of $\mu$ and $n$.


Finally, we also calculate our uncertainty estimates using the maximum likelihood estimation:
\begin{equation}
    \hat{\epsilon}^2 = \frac{1}{\frac{1}{\sigma_e^2}+\frac{1}{\sigma_o^2}},
    \label{eq:error}
\end{equation}
where $\{e,o\}$ are the interleaved even--odd sets and $\sigma^2$ is the sampling-map-weighted variance map for a given even--odd set.

These analytical uncertainty estimates are propagated through the analysis to create our uncertainty on the cross power spectrum. We assume that there are no cross correlations in the noise; each noise pixel in $uv$-space is assumed to be independent. Various techniques, including spatial/frequency windowing, increases pixel-to-pixel correlation. We investigate the strength of the correlations by comparing the analytic noise propagation to the observed noise in the power spectrum in \S\ref{subsec:2D}. 

\subsection{Transforming frequency to $k$-space}
\label{subsec:tTransforming_frequency_to_k-space}

We now have a variety of cubes in $\{k_x,k_y,f\}$-space. In order to go to power spectrum space, we must perform a spectral transform in the frequency direction to go from $f$ to $k_z$. Due to the nature of the data, this includes several steps.

While the binned data are regularly spaced as a function of frequency, the sampling distribution is not constant. We have flagged channels due to RFI in \S\ref{subsec:flagging} and due to channaliser aliasing in \S\ref{subsec:calibration}. The sampling of the $uv$-plane is also inherently non-regular as a function of frequency due to baseline length evolving with frequency. As a result, the $sin$ and $cos$ basis functions of the Fourier transform are not orthogonal to the noise distribution on our sampling; the noise is not independent in the $sin$ and $cos$ basis. However, we can find a basis which is orthogonal. 

We use the Lomb-Scargle periodogram to find this basis \citep{lomb_least-squares_1976,scargle_studies_1982}. A rotation phase is found for each spectral mode that creates orthogonal $cos$-like and $sin$-like eigenfunctions in the noise distribution. This periodogram effectively erases the phase of the data, though this is suitable given that our final goal is to create a power spectrum (a naturally phase-less product).  

The spectral transformation of a relatively small, finite set of data will cause leakage. We mitigate this by multiplying the data by a Blackman-Harris window function in frequency. This will decrease the leakage by about 70\,dB at the first sidelobe, but at the cost of half the effective bandwidth. 

The uneven sampling as a function of frequency can also cause leakage from the bright foregrounds into higher $k_z$. To reduce leakage from the uneven sampling and from the finite spectral transform window, we remove the mean of the data before applying the window function. 
To preserve the DC term in the power spectrum, we add this value back to the zeroth mode after our spectral transform. This preserves power while improving dynamic range. 

We do not expect this mean subtraction and reinsertion to affect our measured EoR power spectrum because the vast majority of the DC-mode power is from foregrounds. In addition, the zeroth $k_z$-mode is not included in our EoR estimates. All our signal loss simulations include this technique to verify that there is no unexpected effect on higher $k_z$-modes.

The periodogram dual, $\eta$, is related to wavenumber space through
\begin{equation}
    k_z \approx \frac{2 \pi H_0 f_{21} E(z)}{c(1+z)^2} \eta,
\end{equation}
where $c$ is the speed of light, $z$ is redshift, $H_0$ is the Hubble constant in the present epoch, $f_{21}$ is the frequency of the 21\,cm emission line, $E(z)$ describes how $H_0$ evolves as a function of redshift, and $k_z$ is the wavenumber along the line-of-sight. This is an approximation; it is valid for all reasonable parameters and avoids a frequency kernel \citep{morales_toward_2004}.

We have chosen our power estimator to be along $k_z$ rather than along the time delay of the electric field propagation between elements $k_\tau$. This subsequently determines \textit{how} the foreground contaminates our final power spectrum \citep{morales_understanding_2018}. Given our choice of basis, we are constructing an \textit{imaging} (or reconstructed sky) power spectrum.

\subsection{Power spectrum products}
\label{subsec:Various_power_spectrum_products}

We now have the power spectra estimates for the mean, the noise, and the uncertainty estimates as a function of $\{k_x,k_y,k_z\}$.

We construct our power spectra by subtracting the power of the even--odd difference from the power of the even--odd summation, and dividing by 4:
\begin{equation}
    p = \frac{p_{\hat{\mu}} - p_{\hat{n}}}{4} = \frac{\hat{\mu}^2 - \hat{n}^2}{4} = \frac{\frac{x_e}{\sigma_e^2} \frac{x_o}{\sigma_o^2}}{\frac{1}{\sigma_e^2}+\frac{1}{\sigma_o^2}}.
\end{equation}
This gives us the same result as the cross power between the even and odd cubes, and
the additional products also allow us to build more diagnostics and to carry the noise throughout the pipeline.

We would like to perform averages over these cubes to generate the best possible limits and to generate diagnostics. To do so, we must assume that the EoR is spatially homogeneous and isotropic, which allows for spherical averaging in Fourier space \citep{morales_toward_2004}. Much like the creation of even--odd sum and difference cubes, we perform the weighted average of pixels. The only difference is that the variances $\sigma^2$ are no longer Gaussian, but rather Erlang variances which can be propagated from the original variances.

We can now create various power spectrum products in 2D and 1D with our 3D power spectrum cube. Uncertainty estimates, measured noise contributions, and expectation values that we have generated in {\eppsilon} will also be used in our diagnostics.

\section{Power spectrum diagnostics}
\label{sec:power_spectrum_diagnostics}

We create a variety of diagnostic power spectrum plots using various averaging schemes. These are essential to understanding contributions to the power spectrum, and are vital to assessing changes to the analysis. In this section, we use MWA Phase I data as a specific example, however all of these diagnostic plots are part of any analysis with FHD/{\eppsilon}.

\subsection{2D power spectrum}
\label{subsec:2D}

\begin{figure}[t]
\centering
	\includegraphics[width = .3\textwidth]{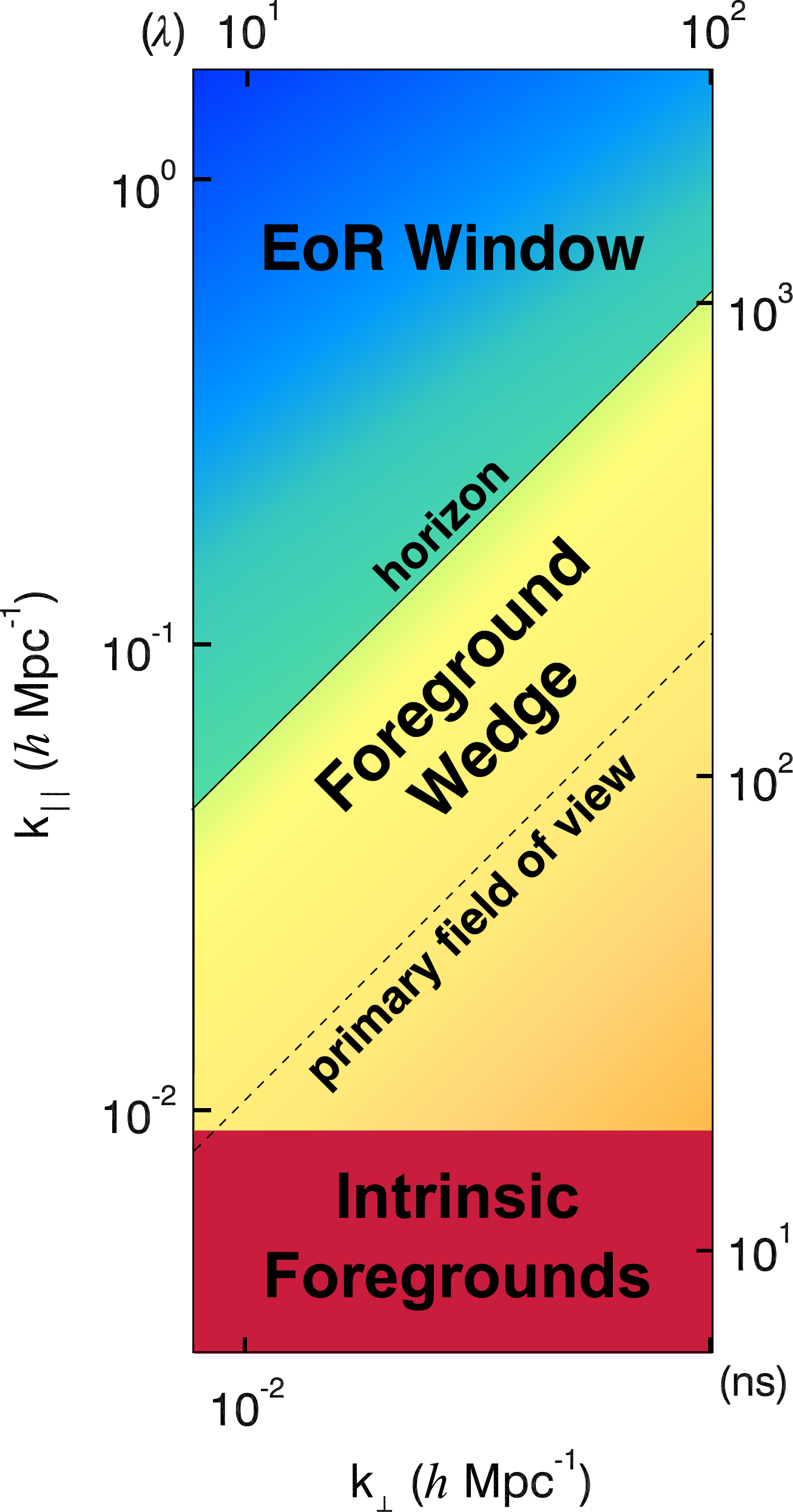}
	\caption[Schematic representation of a 2D power spectrum]{A schematic representation of a 2D power spectrum. Intrinsic foregrounds dominate low $k_{\parallel}$ (modes along the line-of-sight) for all $k_{\bot}$ (modes perpendicular to the line-of-sight) due to their relatively smooth spectral structure. Chromaticity of the instrument mixes foreground modes up into the foreground wedge. The primary-field-of-view line and the horizon line are contamination limits dependent on how far off-axis sources are on the sky. Foreground-free measurement modes are expected to be in the EoR window.}
	\label{fig:cartoon}
\end{figure}

The most useful diagnostic we have is the 2D power spectrum. We average our $\{k_{x},k_{y},k_{z}\}$-power measurements along only the angular wavenumbers $\{k_{x},k_{y}\}$ in cylindrical shells. The resulting power spectrum is a function of modes perpendicular to the line-of-sight ($k_\bot$) and modes parallel to the line-of-sight ($k_\parallel$) shown in Figure~\ref{fig:cartoon}. The $\{k_\parallel,k_\bot\}$-axes have been converted into $\{\tau,\lambda\}$ on the right and top axes---delay in nanoseconds and baseline length in wavelengths, respectively.

Wavenumber space is crucial for statistical measurements due to the spectral characteristics of the foregrounds. Diffuse synchrotron emission and bright radio sources, while distributed across the sky, vary smoothly in frequency (e.g.\ \citet{matteo_radio_2002,peng_oh_foregrounds_2003}). Only small $k_\parallel$-values are theoretically contaminated by bright, spectrally smooth astrophysical foregrounds. Since the foreground power is restricted to only a few low $k_\parallel$-modes, larger $k_\parallel$-values tend to be free of these intrinsic foregrounds in wavenumber space.

However, interferometers are naturally chromatic. This chromaticity distributes foreground power into a distinctive foreground wedge due to the mode-mixing of power from small $k_\parallel$-values into larger $k_\parallel$-values as illustrated in Figure~\ref{fig:cartoon} \citep{datta_bright_2010,morales_four_2012,vedantham_imaging_2012,parsons_per-baseline_2012,trott_impact_2012,hazelton_fundamental_2013,thyagarajan_study_2013,pober_opening_2013,liu_epoch_2014}. The primary field-of-view line and the horizon line are the expected contamination limits caused by measured sources in the primary field-of-view and the sidelobes, respectively. The remaining region, called the EoR window, is expected to be contaminant-free. Because the power of the EoR signal decreases with increasing $k_{||}$, the most sensitive measurements are expected to be in the lower, left-hand corner of the EoR window.

\begin{figure*}
\centering
	\includegraphics[width = \textwidth]{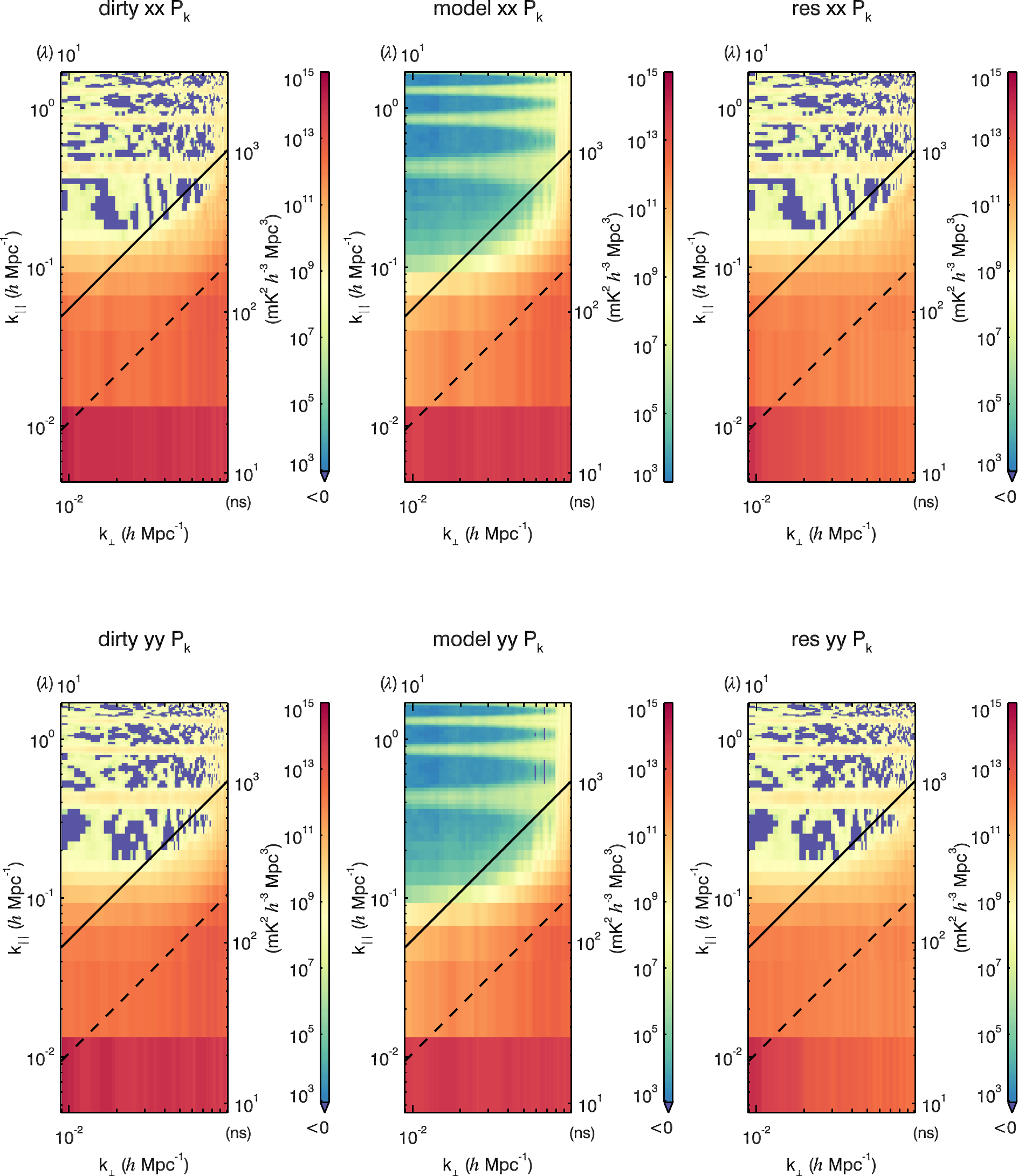}
	\caption[2D power spectra for the calibrated data, model, and residual from 8/23/2013]{The 2D power spectra for the calibrated data, model, and residual for polarisations $XX$ and $YY$ of an integration of 64 MWA observations ($\sim$2\,hrs of data) from August 23, 2013. The characteristic locations of contamination are very similar to Figure~\ref{fig:cartoon}, with the addition of contamination at $k_\parallel$ harmonics due to flagged frequencies with channeliser aliasing. Voxels that are negative due to thermal noise  are dark purple-blue.}
	\label{fig:ex_data_panel}
\end{figure*}

The intrinsic foregrounds, the foreground wedge, the primary field-of-view line, the horizon line, and the resulting EoR window all have characteristic shapes in 2D power spectrum space. Thus, it is a very useful diagnostic space for identifying contamination in real data. We generate a 2D power spectrum as a function of $k_\bot$ and $k_\parallel$ for each of the calibrated data, model, and residual data sets in polarisations $pp = XX$ and $qq = YY$. Shown in Figure~\ref{fig:ex_data_panel} is an example 2D power spectrum panel from a MWA integration of August 23, 2013.

We use this 2D power spectrum panel to help determine if expected contamination occurred in expected regions. In addition to the features shown in Figure~\ref{fig:cartoon}, there is harmonic $k_\parallel$ contamination in the EoR window which is constant in $k_\bot$. This is caused from regular flagging of aliased frequency channels due to the polyphase filter banks in the MWA hardware, which creates harmonics in $k$-space.

\begin{figure*}
\centering
	\includegraphics[width = .7\textwidth]{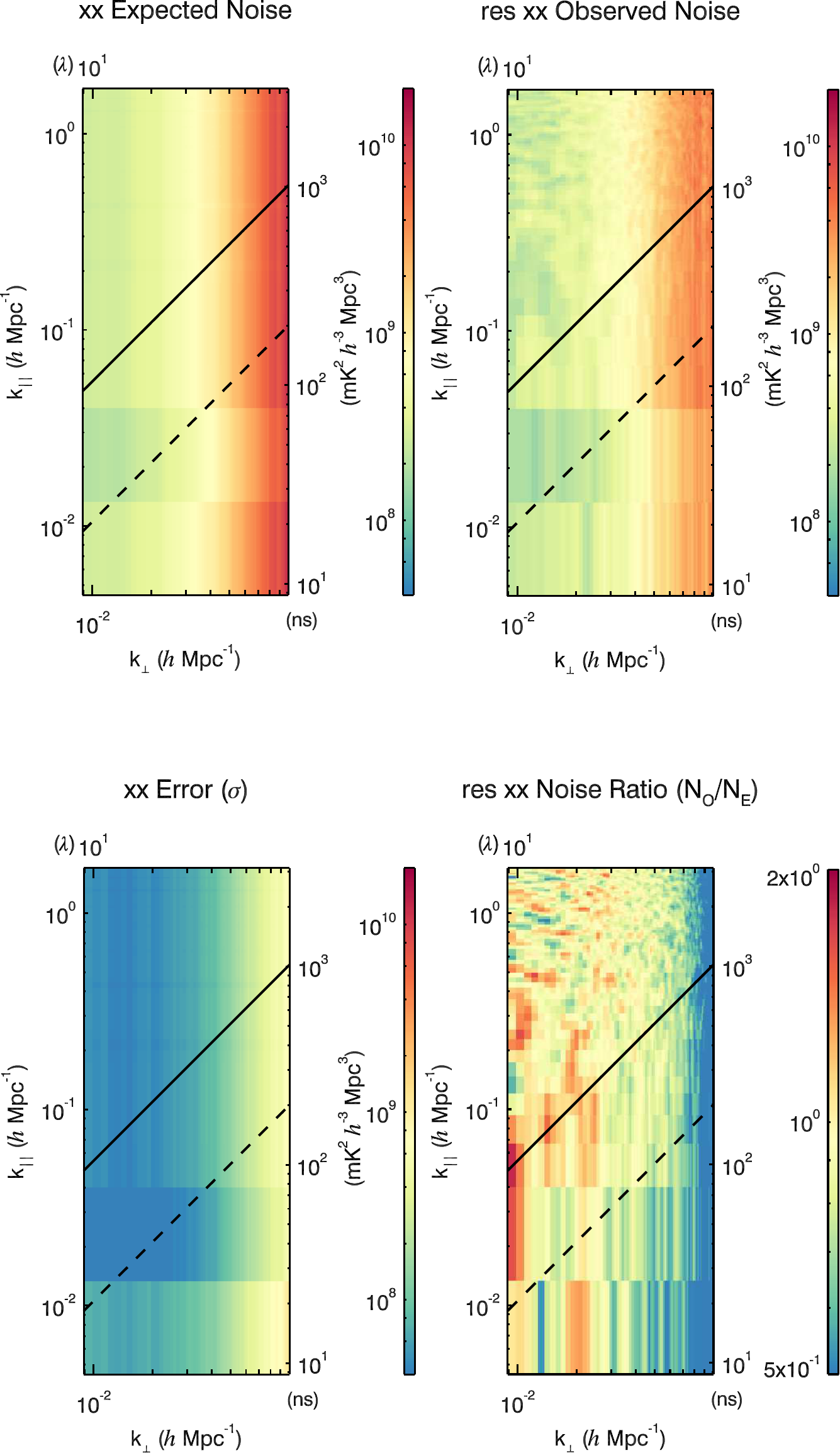}
	\caption[2D power spectra for expected noise, observed noise, error bars, and noise ratio from 8/23/2013]{The 2D power spectra for expected noise, observed noise, error bars, and noise ratio of an integration of 64 observations from August 23, 2013 in instrumental $XX$ for the MWA. The observed noise ($\mathrm{N_O}$) and expected analytically propagated noise ($\mathrm{N_E}$) have a ratio near 1, indicating our assumptions are satisfactory. The error bars are related to the observed noise via Equation~\ref{eq:noises}.}
	\label{fig:ex_noise_panel}
\end{figure*}

In order to verify our error propagation, we also calculate observed and expected 2D noise power spectra. The expected noise, observed noise, propagated error and noise ratio 2D power spectra are shown in Figure~\ref{fig:ex_noise_panel}. The observed noise is the power spectrum of the maximum likelihood noise in Equation~\ref{eq:ml}, so it is the realisation of the noise in the data. The expected noise and error are the expectation value and the square root of the variance, respectively, of the analytically propagated error distributions. We calculate these by propagating the initial Gaussian distributions through the full analysis:
\begin{equation}
    \textrm{Var}[N] = \frac{1}{\sum\limits_{i=0}^n\frac{1}{4\sigma_i^4}} ~~ ~~ ~~ \textrm{E}[N] = \frac{\sum\limits_{i=0}^n\frac{1}{2\sigma_i^2} }{\sum\limits_{i=0}^n\frac{1}{4\sigma_i^4}},
    \label{eq:noises}
\end{equation}
where $\textrm{Var}[N]$ is the variance on the noise, $\textrm{E}[N]$ is the expected noise, $n$ is the number of pixels in the average, and $\sigma^2$ is the original Gaussian variance. 

The final plot in Figure~\ref{fig:ex_noise_panel} is the ratio of our observed noise to our expected noise. This investigates our assumption of uncorrelated pixels in $uv$-space during noise propagation. There will be fluctuations given different noise realisations on the observed noise, but the ratio should fluctuate around one. In practice, this is approximately true; the mean of the noise ratio from 10 to 50$\,\lambda$ is $\sim$0.9. The propagated noise is therefore slightly higher, indicating that we overestimate our noise due to assuming a lack of correlation.

\subsection{1D power spectrum}

Averaging to a 1D power spectrum harnesses as much data as possible, thereby making the limits with the lowest noise. However, the 1D power spectrum also has the ability to be an excellent secondary diagnostic after the 2D power spectrum. While characteristic locations of contamination are easier to distinguish on 2D plots, 1D diagnostics are more able to distinguish subtleties.

\begin{figure*}
\centering
	\includegraphics[width = \textwidth]{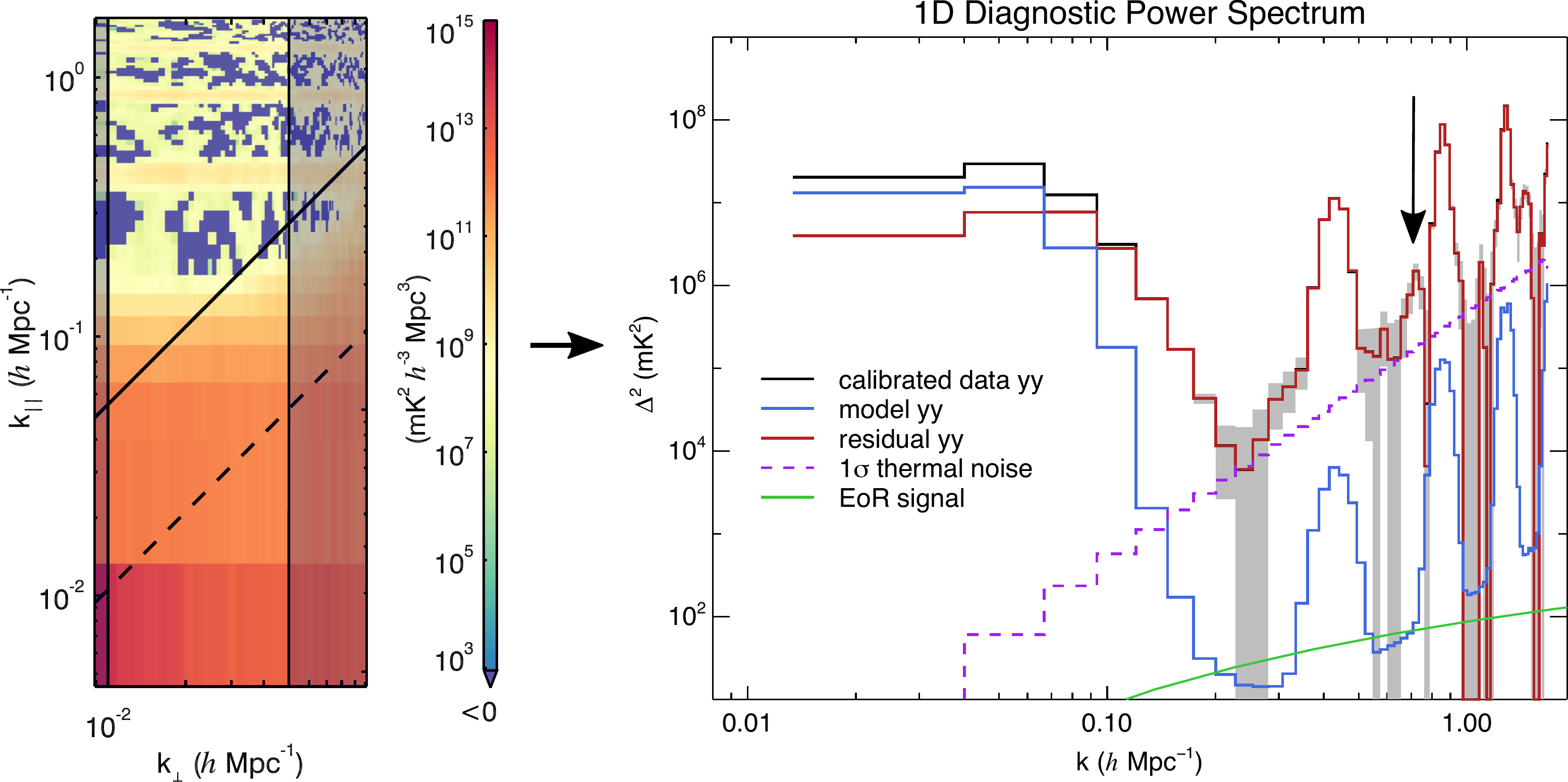}
    	\caption[1D power spectra for calibrated data, model, residual, theoretical EoR, and thermal noise contribution from 8/23/2013]{1D power spectra as a function of $k$ for calibrated data (black), model (blue), residual (red), 2$\sigma$ uncertainties (shaded grey), theoretical EoR (for comparison, green), and the thermal noise (dashed purple) of an integration of 64 MWA observations from August 23, 2013 for instrumental $YY$. The 2D power spectrum highlights the bins that went into the 1D averaging, which we can modify to exclude the foreground wedge when making limits. A cable-reflection contamination feature at $0.7$\,\textit{h}\,Mpc$^{-1}$ is more obvious in this 1D power spectrum, which highlights the importance of using 1D space as a secondary diagnostic.}
	\label{fig:1Ddiagnostic}
\end{figure*}

The most simplistic 1D power spectrum is an average in spherical shells of all voxels in the 3D power spectrum cube which surpass a low-weight cutoff (see Appendix~\ref{appendix} for more details). This includes all areas of contamination explored in \S\ref{subsec:2D} which can obscure low-power regions. Therefore, a typical secondary diagnostic is a 1D power spectrum generated only from pixels which fall within 10 and 50\,$\lambda$ in $k_\bot$, shown in Figure~\ref{fig:1Ddiagnostic}. This will include intrinsic foregrounds and some of the foreground wedge, but will avoid contaminated $k_\bot$-modes with poor $uv$-coverage for the MWA. The exact $k_\bot$-region for this diagnostic will depend on the instrument.

The typical characteristic contamination shapes are present in Figure~\ref{fig:1Ddiagnostic}, like the intrinsic foregrounds, foreground wedge, and flagging harmonics, all of which are expected given the 2D power spectra. However, a new contamination feature occurs at about $0.7$\,\textit{h}\,Mpc$^{-1}$. This contamination is from a cable reflection in the elements with 150\,m cables. It was significantly reduced due to the calibration technique in \S\ref{cableref}, but spectral structure from unmodelled sources limit the precision \citep{barry_calibration_2016}. Much like with the aliased channel flagging, spectrally repetitive signals will appear as a bright contamination along $k_\bot$, translating to a near-constant 1D contribution if high in $k_\parallel$.

This highlights the potential of utilizing multiple types of power spectra. The 2D power spectrum is a powerful diagnostic due to characteristic locations of contamination. However, it is useful to also use 1D diagnostic power spectra to more quantitatively measure contamination, which helps in discerning smaller contributions.

The cable reflections and flagged aliased channels lead to bright contamination due to their modulation as a function of frequency. This is an important revelation; we must minimise all forms of spectrally repetitive signals during power spectrum analysis. If a spectral mode is introduced in the instrument or in the processing, that mode cannot be used to detect the EoR.

\subsection{2D difference power spectrum}
\label{subsec:2D_difference}

Often, we would like to compare a new data analysis technique to a standard to assess potential improvements. We can do this with a side-by-side comparison of 2D power spectra, but are limited by the large dynamic range of the color bar. Instead, we create 2D difference power spectra.

We take a 3D bin-by-bin difference between two $\{k_{x},k_{y},k_{z}\}$-cubes and then average to generate a 2D difference power spectrum. The reference or standard is subtracted \textit{from} the new run, and the 2D difference power spectrum varies positive and negative depending on power levels. We choose a red--blue log-symmetric color bar to indicate sign, where red indicates an increase in power relative to the reference and blue indicates a decrease in power.

\begin{figure*}
\centering
	\includegraphics[width=\textwidth]{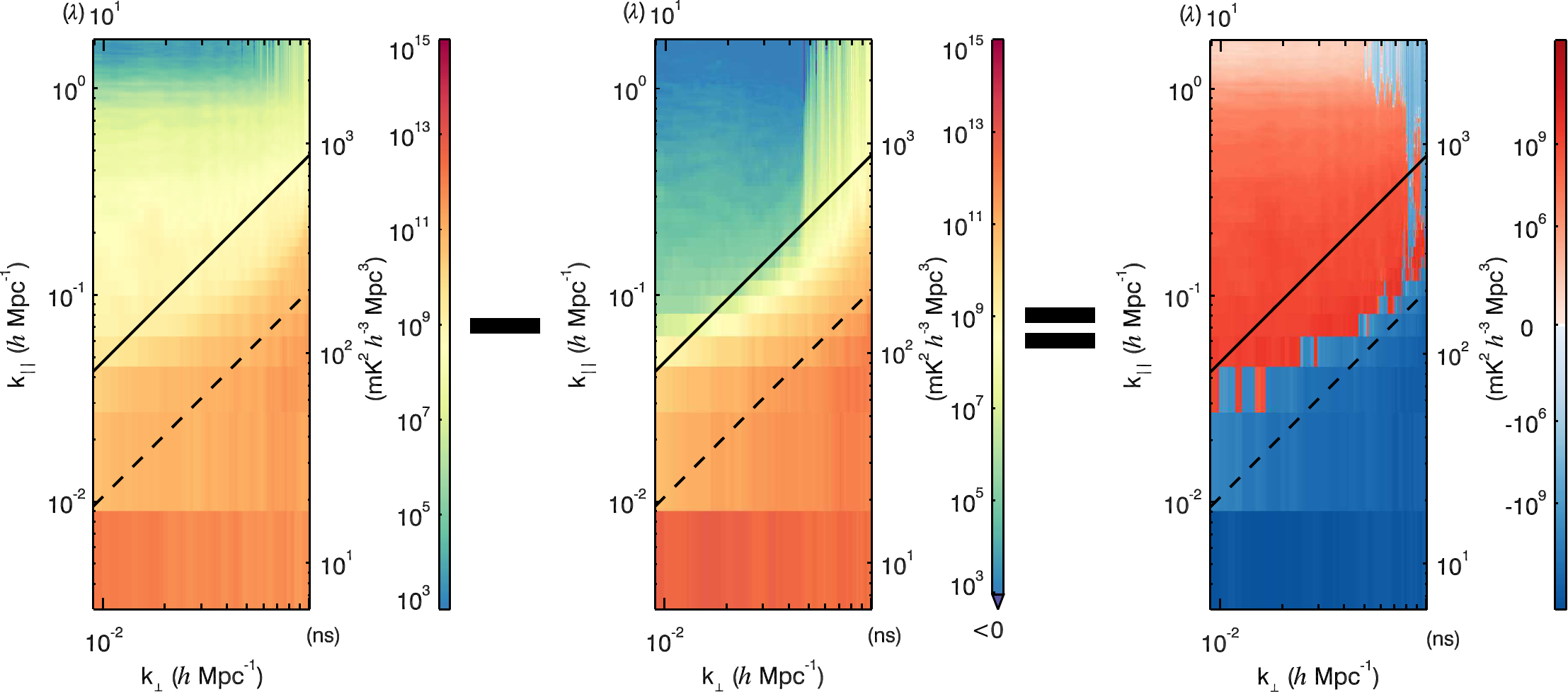}
	\caption[2D difference power spectrum example]{The subtraction of a residual 2D power spectrum (left) and a reference residual 2D power spectrum (middle) to create a difference 2D power spectrum (right). Red indicates a relative excess of power, and blue indicates a relative depression of power. This diagnostic is helpful in determining differences in plots that inherently cover twelve orders of magnitude.}
	\label{fig:diff}
\end{figure*}

Figure~\ref{fig:diff} shows an example of a 2D difference power spectra. The new data analysis run is the left panel, the reference is the middle panel, and the 2D difference is the right panel. For this example, we have chosen a new data analysis run that had excess power in the window and a decrease in power in the foreground wedge compared to the reference. In general, we make 2D difference power spectra for dirty, model, and residual, for both $XX$ and $YY$ polarisations to match the six-panel plot in Figure~\ref{fig:ex_data_panel}.

\section{Signal loss simulation}
\label{subsec:signal_loss}

The FHD/{\eppsilon} pipeline can be used as an \textit{in situ} simulation to test for signal loss. This validates our final power spectra results, allowing for confidence in our EoR upper limits. 

FHD is an instrument simulator by design; we create model visibilities from all reliable point sources and realistic beam kernels. We can use these model visibilities as the base of an input simulation of the instrument and sky. A statistical Gaussian EoR is Fourier-transformed to $uv$-space, encoded with instrumental effects via the $uv$-beam, and added to point source visibilities to create data that we can test for signal loss. These \textit{in situ} simulation visibilities are input into the FHD/{\eppsilon} pipeline. They are treated like real data, and are subject to all real data analyses.

We perform four \textit{in situ} simulations to validate our pipeline:
\begin{itemize}
  \item Compare input EoR signal to output EoR signal with no additional foregrounds or calibration. This simple test demonstrates consistency in power spectrum normalisation and signal preservation.
  \item Allow all sources to be used in the calibration and subtraction model and compare the output power to the input EoR signal. By recovering the input EoR signal, we demonstrate that the addition of foregrounds and calibration does not result in signal loss.
  \item Use an imperfect model for calibration and subtraction and compare the output power to the input EoR signal. Particularly, we calibrate with a global bandpass (Equation \ref{eq:global_bp}) in the comparison. The calibration errors caused by spectral structure from unmodelled sources cause contamination in the EoR window (see \citet{barry_calibration_2016}), but there is no evidence of signal loss.
  \item Use an imperfect model for subtraction but a perfect model for calibration, and compare the output power in the EoR window to the input EoR signal. We can recover the EoR signal in some of the EoR window, demonstrating that we can theoretically detect the EoR even with unsubtracted foregrounds.
\end{itemize}

\begin{figure}[h]
\centering
	\includegraphics[width=\columnwidth]{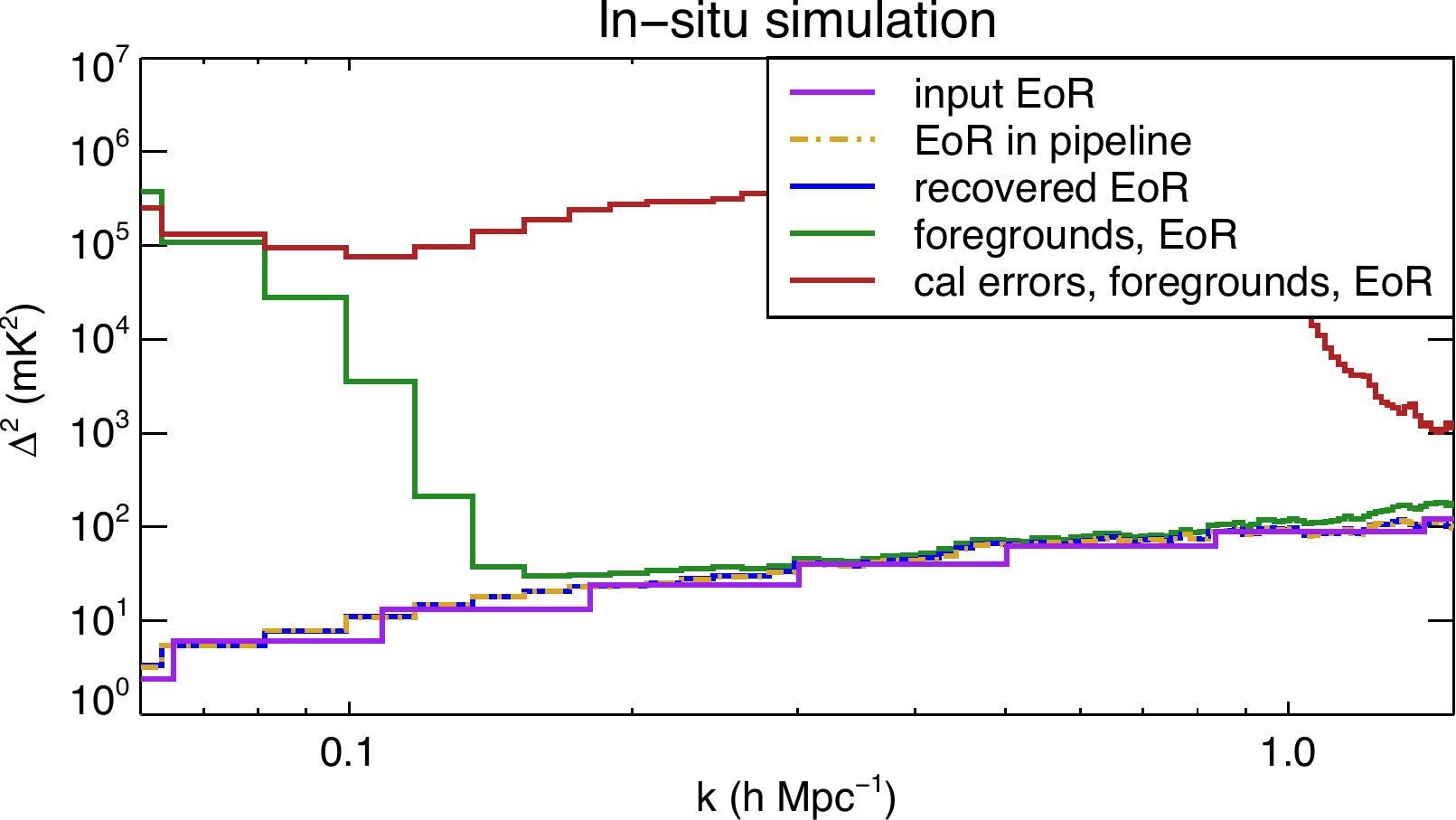}
	\caption[Recovered EoR]{The signal loss simulations of the FHD/{\eppsilon} pipeline. We recover the input EoR signal (purple) for most $k$ modes if: 1) an EoR signal is propagated through the pipeline without foregrounds or calibration (orange, dashed), 2) all foregrounds are used for calibration and are perfectly subtracted (blue), and 3) all foregrounds are used for calibration but are not perfectly subtracted (green). We do not recover the EoR if there are calibration errors (red), however these errors are not indicative of signal loss.}
	\label{fig:signal_loss}
\end{figure}

Figure~\ref{fig:signal_loss} shows the output power from each test alongside the underlying, input EoR signal. For these simulations, we use the MWA Phase I instrument without channeliser effects. They perform as expected, recovering the EoR signal when possible. However, at high $k$-modes, the foreground simulation without calibration errors is contaminated due to a gridding resolution systematic \citep{beardsley_first_2016}.

If regular flagging of aliased channeliser effects are included, more modes are contaminated in the EoR window, as seen in the model data of Figure~\ref{fig:1Ddiagnostic}. This is a consequence of the lack of inverse-variance weighting during the spectral transform. Implementing a more complex weighting scheme and/or modifying the actual instrument will be necessary to recover those modes in the future. 

It is important for an EoR power spectrum analysis to characterise potential signal loss in order for EoR limits to be valid \citep{cheng_characterizing_2018}. These tests demonstrate that we are not subject to signal loss within our pipeline. None of our analysis methodologies, from FHD all the way through {\eppsilon}, are artificially removing the EoR signal. In addition, these simulations can be expanded to include other effects in the future, such as channeliser structure, diffuse emission, and beam errors.

These signal loss simulations can only test \textit{relative} forms of signal loss. For example, this cannot test the validity of the initial encoded instrument effects on the input (i.e the \textit{in situ} simulation or EoR visibilities), nor can it test the accuracy of the sky temperature used in normalisation. 



\section{Overview}

We have built an analysis framework that takes large quantities of measured visibilities and generates images, power spectra, and other diagnostics. Four modularised components exist: pre-analysis flagging and averaging, calibration and imaging, integration, and error-propagated power spectrum calculations. The accuracy of each component is crucial due to the level of precision needed in an EoR experiment. In particular, the analysis handled by FHD and {\eppsilon} is complicated and multifaceted, necessitating constant refinement and development to ensure accuracy, precision, and reproducibility.

Many data products are produced with our analysis pipeline. We make instrumental polarisation images for each observation for calibrated, model, and residual data. As for power spectra, we make 2D and 1D representations for the calibrated, model, and residual data from integrated cubes along with propagated noise and propagated uncertainty estimates. These form the basic outputs within the FHD/{\eppsilon} pipeline. 

All examples provided have been from MWA Phase I data, however FHD/{\eppsilon} is more generally applicable. It has been used to publish MWA Phase II data \citep{li_comparing_2018} and PAPER data \citep{kerrigan_improved_2018}, with plans to analyse HERA data. Flexibility within the pipeline allows for constant cross-validation and a wide variety of use cases. 

It is crucial to have a fully verifiable pipeline, and this has been the major motivation behind FHD/{\eppsilon}. All of our diagnostic outputs, error propagation products, and signal loss simulations are a part of a signal preservation narrative. By establishing confidence in our data analysis, we can pursue and publish credible EoR upper limits and measurements.

The FHD/{\eppsilon} pipeline is readily available online, and developed fully in public. Anyone can view our progress and critically analyse our methodologies. This is a resource for the community; FHD/{\eppsilon} is an open-source example of the necessary precision techniques required for EoR power spectrum analysis.


\begin{acknowledgements}
This research was supported by the Australian Research Council Centre of Excellence for All Sky Astrophysics in 3 Dimensions (ASTRO 3D), through project number CE170100013. This work has been funded by National Science Foundation grants AST-1410484, AST-1506024, AST-1613040, AST-1613855, and AST-1643011. APB is supported by an NSF Astronomy and Astrophysics Postdoctoral Fellowship under \#1701440.
\end{acknowledgements}

\begin{appendix}

\section{Power spectrum volume normalization}
\label{appendix}

The data products handed from FHD to {\eppsilon} are reconstructions of the apparent sky---a 1\,Jy source at the half-power point of the beam will appear as a $\frac{1}{2}$\,Jy source. A volume factor is needed to convert the power spectrum of the apparent image into a properly normalised cosmological power spectrum. The normalisation factor needed for an area of the $uv$-plane estimated from an isolated baseline differs from the normalisation factor for an area estimated from many overlapping baselines. Formally, this is related to the covariance of the overlapping visibilities, as described in \citet{liu_epoch_2014}. Carrying the full covariance through the pipeline is computationally intractable, so we separately normalise regions with dense and poor $uv$-coverage.

\begin{figure}
\centering
	\includegraphics[width=\columnwidth]{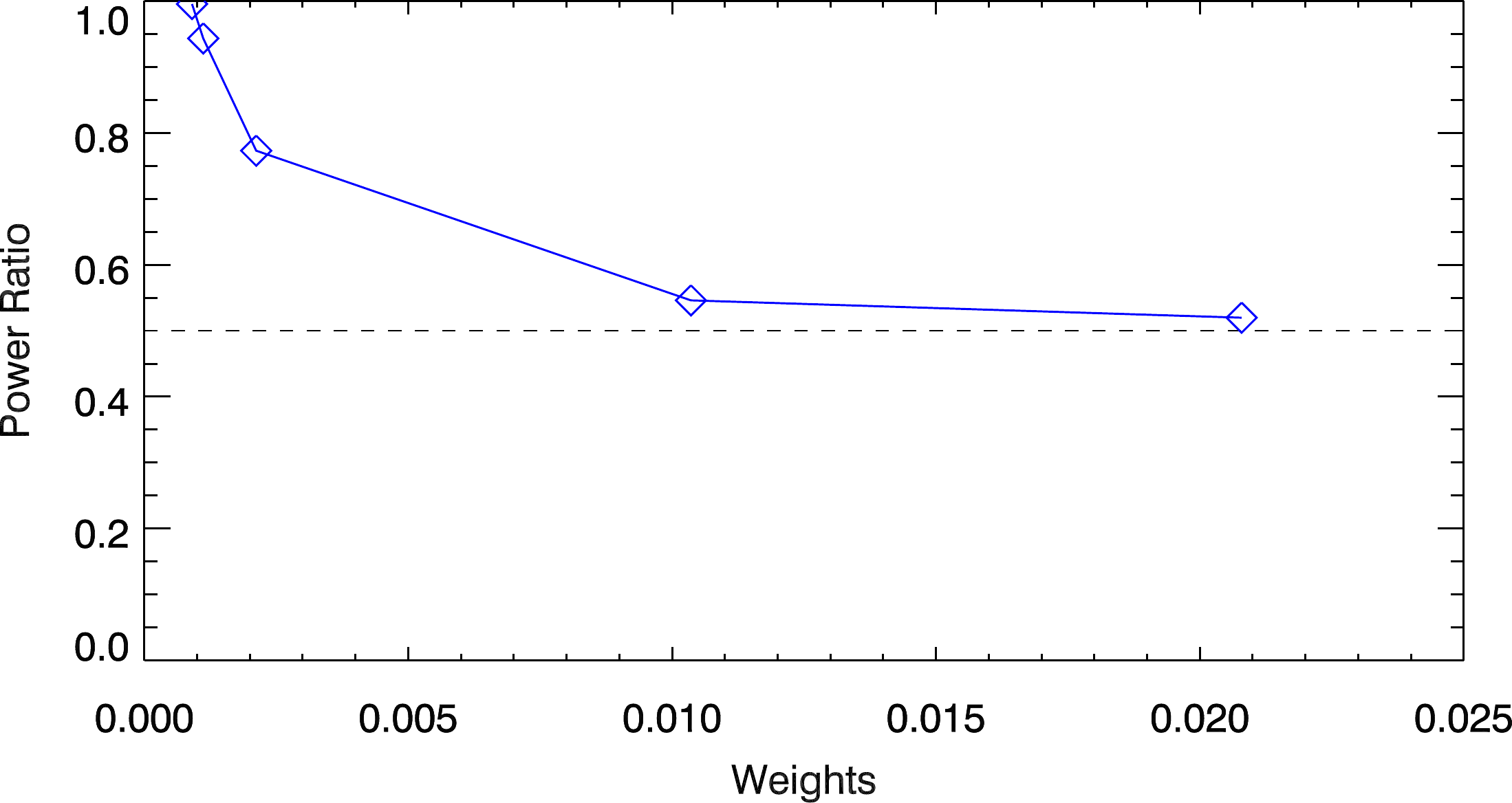}
	\caption[Simulation suite]{Results of our end-to-end simulations through FHD/{\eppsilon} of a flat power spectrum signal. The blue points give the ratio of the reconstructed 1D power spectra using the calculated sparse normalisation to the input power level (which is flat in $k$). The x axis is a measure of baseline density--it gives the average baseline weight gridded to each $uv$-pixel in the simulation. In constructing limits and 1D power spectra, we only use regions of the $uv$-plane where the minimum weight (as a function of frequency) is greater than or equal to 1.}
	\label{fig:sim_suite}
\end{figure}

We demonstrate this effect with end-to-end simulations in FHD/{\eppsilon} of a flat power spectrum signal (constant power as a function of $k$) and randomly located baselines with increasing density in the $uv$-plane. For each simulation, a baseline density is selected and baselines are randomly placed in a $uv$-plane. The corresponding visibilities are simulated for a stochastic signal with a flat power in $k$, and then gridded to calculate a 1D power spectrum. The reconstructed 1D power spectra are flat in $k$, but the normalisation relative to the calculated sparse normalisation depends on baseline density, as shown in Figure~\ref{fig:sim_suite}. In the limit of very sparse baselines, there are almost no overlaps and the sparse normalisation is correct, but as the density increases the normalisation decreases quickly and asymptotes to half the sparse normalisation. This factor of 2 in the denominator can be understood as a doubling of the effective area of the $uv$-plane that contributes to the gridded value at any location in the gridded $uv$-plane.

Formally, there is a transition region where the normalisation is difficult to calculate without fully propagating the covariance. However, virtually all our sensitivity comes from regions of the $uv$-plane with very dense overlap. We do not discard the low-density regions in our 2D power spectra, because 2D spectra are primarily diagnostic in nature and are useful in helping us to identify systematics. However, for EoR upper limits and 1D power spectra we discard low density regions to avoid uncertainty in the normalization. Discarding the sparse areas of the $uv$-plane does not significantly decrease our sensitivity and alleviates the need to fully propagate the covariance matrix. 

\end{appendix}

\bibliographystyle{style-class/pasa-mnras}
\bibliography{bib}

\end{document}